\documentclass[traditabstract,fleqn]{aa}
\usepackage{amsmath}  
\usepackage{graphicx}
\usepackage{natbib}
\usepackage[colorlinks=true,linkcolor=blue,urlcolor=blue,citecolor=blue]{hyperref}
\usepackage{siunitx}

\usepackage{threeparttable}
\usepackage[T1]{fontenc}
\usepackage[utf8]{inputenc}

\makeatletter

\usepackage{txfonts}
\bibpunct{(}{)}{;}{a}{}{,}

\titlerunning{Evolution and final fate of massive post-common-envelope binaries}
\authorrunning{Dandan Wei et al.}
\makeatother
\begin{document}

\title{Evolution and final fate of massive post-common-envelope binaries}
\author{Dandan Wei\inst{\ref{HITS}}\thanks{dandan.wei@hotmail.com},
Fabian R. N. Schneider \inst{\ref{HITS},\ref{ZAH}}\thanks{fabian.schneider@h-its.org}, 
Philipp Podsiadlowski\inst{\ref{Oxford},\ref{HITS}},
Eva Laplace\inst{\ref{HITS}},
Friedrich K. Röpke\inst{\ref{HITS},\ref{ITA}},
\and Marco Vetter\inst{\ref{HITS}}}
\institute{Heidelberger Institut f\"{u}r Theoretische Studien, Schloss-Wolfsbrunnenweg 35, 69118 Heidelberg, Germany\label{HITS}
\and Zentrum f\"{u}r Astronomie der Universit\"{a}t Heidelberg, Astronomisches Rechen-Institut, M\"{o}nchhofstr. 12-14, 69120 Heidelberg, Germany\label{ZAH}
\and University of Oxford, St Edmund Hall, Oxford, OX1 4AR, United Kingdom\label{Oxford}
\and Zentrum f\"{u}r Astronomie der Universit\"{a}t Heidelberg, Institut f\"{u}r Theoretische Astrophysik, Philosophenweg 12, 69120 Heidelberg, Germany\label{ITA}
}
\date{Received xxx / Accepted yyy}
\abstract{
Mergers of neutron stars (NSs) and black holes (BHs) are nowadays observed routinely thanks to gravitational-wave (GW) astronomy. In the isolated binary-evolution channel, a common-envelope (CE) phase of a red supergiant (RSG) and a compact object is crucial to sufficiently shrink the orbit and thereby enable a merger via GW emission. Here, we use the outcomes of two three-dimensional (3D) magneto-hydrodynamic CE simulations of an initially 10.0 solar-mass RSG with a 5.0 solar-mass BH and a 1.4 solar-mass NS, respectively, to explore the further evolution and final fate of the remnant binaries (post-CE binaries). Notably, the 3D simulations reveal that the post-CE binaries are likely surrounded by circumbinary disks (CBDs), which contain substantial mass and angular momentum to influence the subsequent evolution. The binary systems in MESA modelling undergo another phase of mass transfer and we find that most donor stars do not explode in ultra-stripped supernovae (SNe), but rather in Type Ib/c SNe. Without NS kicks, the final orbital configurations of our models with the BH companion are too wide to allow for a compact object merger within a Hubble time. NS kicks are actually required to sufficiently perturb the orbit and thus facilitate a merger via GW emission. Moreover, we explore the influence of CBDs observed in 3D CE simulations on the evolution and final fate of the post-CE binaries. We find that mass accretion from the disk widens the binary orbit, while resonant interactions between the CBD and the binary can shrink the separation and increase the eccentricity of the binary depending on the disk mass and lifetime. Efficient resonant contractions may even enable a BH or NS to merge with the remnant He stars before a second SN explosion, which may be observed as gamma-ray burst-like transients, luminous fast blue optical transients, and Thorne-\.Zytkow objects. For the surviving post-CE binaries, the CBD-binary interactions may significantly increase the GW-induced double compact merger fraction. We conclude that accounting for CBD may be crucial to better understand observed GW mergers.
}
\maketitle
\section{Introduction}\label{sec:introduction}
As one of the major uncertain phases in binary evolution, the common-envelope (CE) interaction plays an important role in shrinking the binary separation, thus leading to the formation of short-orbit binaries \citep{2013A&ARvIvanova, 2023LRCAFritz}. CE phases can produce many astrophysical objects such as cataclysmic variables \citep{1976IAUSPaczynski}, X-ray binaries \citep{1998ApJKalogera,2002ApJPodsiadlowski}, progenitors of stripped-envelope supernovae (SNe) \citep{1992ApJPodsiadlowski}, and Type Ia SNe \citep{2005ApJBelczynski,2016ApJAblimit,2023RAALiu}. In addition, CE evolution also contributes to producing a substantial fraction of short-orbit double compact objects, which merge via gravitational-wave (GW) emission \citep{2017ApJTauris,2018MNRASKruckow,2018MNRASVigna,2020A&ABelczynski,2023A&ALi,2024ResPhLi}. Double compact merger events are now detected regularly \citep{2016PhRvLAbbott,2019PhRvXAbbott, 2021PhRvDAbbott, 2021PhRvXAbbott}. To better understand the observed GW merger events, it is crucial to figure out the configuration of binaries following the CE phase (post-CE binary), and the orbital evolution of the post-CE binary before the formation of double compact objects.

In the progenitor evolution of GW merger events within the framework of isolated binaries, a CE interaction between a red supergiant (RSG) and compact companion takes place when the envelope of the RSG engulfs its companion. The CE receives energy and angular momentum during the inspiral of a compact companion. A CE can be ejected successfully if the available energy during the inspiral phase is substantial enough to unbind the CE, leaving behind a tight post-CE binary that formed from the RSG core and the compact object. An accurate final separation of the post-CE binary is crucial to predict the GW merger rate. However, this is still highly uncertain.

To investigate the binary condition after the CE phase (or post-dynamical inspiral phase), two simple parametric approaches have been proposed: the $\alpha$ formalism \citep{1984ApJWebbink,1988ApJLivio} and $\gamma$ formalism \citep{1992ApJPodsiadlowski,2000A&ANelemans,2023ApJDi} that are based on energy and angular momentum conservation, respectively. The free parameters $\alpha$ and $\gamma$ indicate a fixed fraction of released orbital energy or angular momentum during the CE event that contributes to unbinding the envelope. Although the $\alpha$ formalism is widely applied in population synthesis studies \citep[e.g.][]{2002MNRASHurley}, $\alpha$ is still an uncertain parameter \citep{2004ApJPolitano,2019MNRASIaconi,2022MNRASWilson}. More recently, two-step formalisms have been applied \citep{2021A&AMarchant,2022ApJHirai}, which attempt to combine the spiral-in phase with a more standard stable mass-transfer phase in a self-consistent way.

Multi-dimensional hydrodynamical models have been conducted to give insight into the detailed physics during the dynamical inspiral phase of the CE interaction \citep{1995ApJTerman,2012ApJ2012,2016ApJOhlmann,2019MNRASIaconi, 2020MNRASIaconi, 2020A&ASand,2020arXivLaw-Smith, 2022A&AMoreno, 2023LRCAFritz}. A 3D CE magneto-hydrodynamic simulation of a massive RSG star with a compact companion -- a black hole (BH) or a neutron star (NS) -- investigated by \citet{2022A&AMoreno} demonstrates that the final separation after the CE phase is wider than the simple energy arguments predict. The CE simulations with a lower-mass primary in \citet{2017MNRASIaconi} show a similar outcome. Having a wider orbit makes it impossible for the subsequent double compact objects to merge via GW emission within a Hubble time without other orbital evolution events or kicks imparted on the system by SN explosions. Hence, a better understanding of the processes in post-CE binaries is vital for a variety of phenomena such as GW-induced merger events.

A possible additional process during the post-CE evolution is another mass transfer (MT) episode, because of the re-expansion of the RSG core (see e.g. binary models of stripped helium (He) stars or pure He stars with a compact companion in \citealt{2017ApJTauris, 2020A&ALaplace, 2021A&ASchneider, 2021ApJJiang}). As discussed in \citet{2022A&AMoreno}, in using an analytical approach, binary MT changes the orbital evolution of post-CE binaries. 

Another possible process is the formation of a circumbinary disk (CBD) around the supergiant core and the compact object when gravitationally bound envelope material cools down after the dynamical inspiral phase in a CE event \citep{1998ApJSandquist,2011MNRASKashi,2022MNRASLau,2023A&AGagnier,2023ApJTuna}. Plenty of observational evidence demonstrates the existence of CBDs around low-mass binaries, for example, post-asymptotic giant branch (post-AGB) binaries and post-red giant branch (post-RGB) binaries \citep{2013A&ABujarrabal,2018arXivVan,2022MNRASLi}. Analytical CBD models have been proposed to be responsible for pumping eccentricities in the post-AGB and RGB binaries \citep{1991ApJArtymowicz, 1996ASICLubow,2000prplLubow,2011MNRASKashi,2020A&AOomen,2023MNRASIzzard}, and fast orbital shrinkage of BH X-ray binaries \citep{2018ApJXu,2019ApJChen}. Recently, numerical simulations of CBDs have been conducted to better understand the interaction between the CBD and the inner binary \citep{2016ApJMunoz,2019ApJMunoz,2020ApJMunoz,2022MNRASDittmann,2023MNRASSiweka,2023MNRASSiwekb}. \citet{2023MNRASSiwekb} show that CBDs cannot only result in contraction but also expansion of the binary orbit depending on the binary configurations. The 3D CE simulations in \citet{2022A&AMoreno} and Vetter et al. (in prep) show that the post-CE binaries are likely surrounded by CBDs, which contain substantial gravitationally bound material and a significant amount of angular momentum. \citet{2011MNRASKashi} were the first to point out the importance of such post-CE CBDs, as they can strongly affect the post-CE evolution of the surviving binaries. In order to reliably predict the final fate of such systems, an extensive study of the interaction between the CBD and the inner binary seems inevitable. In addition, the subsequent SN explosion is expected to perturb the binary orbit, which can greatly affect predictions of GW merger rates.

In this work, we investigate the evolution and final fate of massive post-CE binaries by following the MT using MESA and analysing the CBD models analytically based on the outcomes of 3D CE simulations. In Sect.\,\ref{sec:methods}, we describe the initial setup of the post-CE binary and the physical assumptions of the binary evolution calculations in MESA. We further provide the methodology of the analytical CBD models. In Sect.\,\ref{Additional mass transfer}, we demonstrate how the binary interaction affects the evolution of post-CE binaries with different compact objects, and the ultimate double compact merger fraction within a Hubble time. The influence of a CBD-binary interaction on the evolution and final fate of post-CE binaries is shown in Sect.\,\ref{CBD-binary interactions}. We discuss and conclude our results in Sect.\,\ref{Discussion} and Sect.\,\ref{Concusion}, respectively.

\section{Methods}\label{sec:methods}
In this section, we introduce the outcome of 3D CE simulations and then explain how we construct the post-CE binary configuration based on these 3D CE simulation outcomes. The physical assumptions of the binary evolution calculations in MESA are described. In addition, we provide the methodology of the analytical CBD models, which includes the accretion from the CBD and the resonant interactions.

\subsection{Post-CE Arepo model}
\label{Arepo_method}

\begin{figure}
\begin{center}
\includegraphics[width=0.5\textwidth]{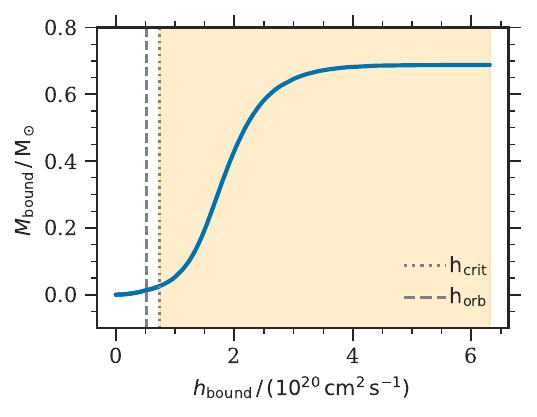}
\caption{Accumulated mass of the bound material as a function of the specific angular momentum at the end of the 3D CE simulation in the case of a BH companion. The vertical dotted and dashed grey lines indicate the critical- and orbital-specific angular momentum, respectively. The yellow shaded area indicates the material with the specific angular momentum larger than the critical value.}
\label{fig:M_J_bound_material}
   \end{center}
\end{figure}

Three-dimensional CE magneto-hydrodynamic simulations of an initially 10\,$\mathit M_\odot$ RSG after core helium burning have been conducted in \citet{2022A&AMoreno} with both a BH and a NS companion. Vetter et al. (in prep) extended these 3D CE simulations by approximately a factor of $\sim 4$ in time until $90\,\%$ of the former envelope mass is ejected. To mitigate the increasing computational costs in the late stages of the CE interactions caused by the progressively decreasing cell sizes within the gravitational softening lengths of the primary star core and companion \citep[see e.g.][]{2017A&AOhlmann}, Vetter et al. (in prep) utilised a new refinement approach. In this approach, the physical resolution inside the gravitationally softened potentials is kept constant throughout the CE interactions. According to the outcomes of Vetter et al. (in prep), the remnant RSG core has mass of 2.97\,$\mathit M_\odot$ and the orbital separation of the post-CE binary with a BH companion is 37\,$\mathit R_\odot$, while it is closer for the NS companion (14\,$\mathit R_\odot$) because of its more pronounced inspiral. At the end of 3D CE simulations, eccentric orbits are found in both cases, with eccentricities of $e=0.06$ for the BH companion and $e=0.01$ for the NS companion.

At the end of both CE simulations, gravitationally bound material is observed in \citet{2022A&AMoreno} and Vetter et al. (in prep). A hot and thick CBD is thought to form initially after the dynamical plunge-in phase during the CE interactions. The disk then cools radiatively on a short timescale, only a few orders of magnitude longer than the dynamical time, eventually leading to a thin CBD \citep{2023ApJTuna}. The cooling timescale is approximately several hundred years if we apply Eq.\,(17) in \citet{2020MNRASSchneider} to estimate how fast the releasing gravitational potential energy is lost by radiatively cooling. Given the shorter cooling timescale than the remaining lifetime of the remnant RSG core, a thin CBD can form and interact with the inner binary. The mass of the bound material in the case of a BH companion is approximately 0.69\,$\mathit M_\odot$, as shown in Fig.\,\ref{fig:M_J_bound_material}. Nearly 97\% of the bound material has a larger specific angular momentum than the critical value needed to form a CBD. The critical value of the specific angular momentum $h_{\rm crit}$ is given by the inner edge of the CBD, 
\begin{equation} 
\label{eqs.1}
h_{\rm crit} = \sqrt{2\rm G \mathit{\left(M_{\rm 1} + M_{\rm 2}\right)a}}=\sqrt{2}\,h_{\rm orb}. 
\end{equation}
For comparison, we provide the orbital-specific angular momentum $h_{\rm orb}$. Here, $M_{\rm 1}$, $M_{\rm 2}$ and $a$ are the mass of the two components, and the orbital separation, respectively. The inner edge of the CBD is assumed to be located at $2a$, according to the smoothed particle hydrodynamics (SPH) simulations of \citet{1994ApJArtymowicz} and \citet{2000prplLubow}.  Notably, the total angular momentum of nearly 1.4 times the orbital angular momentum of the post-CE binary is carried by the bound mass. A similar ratio regarding the total angular momentum exists in the case of a NS companion, even though less mass (${\sim}\, 0.17\,\mathit M_\odot$) is found to be bound surrounding the inner binary. Having such a substantial amount of angular momentum and mass makes it possible for the CBD to alter the evolution and final fate of the post-CE binaries \citep[cf.][]{2011MNRASKashi}.

\subsection{Post-CE binary configuration}
\label{post-CE_method}

\begin{figure}
\begin{center}
\includegraphics[width=0.5\textwidth]{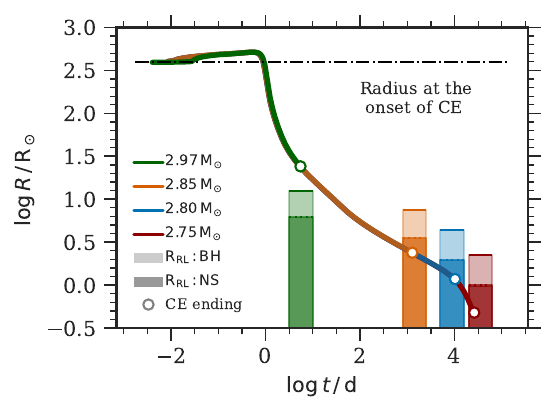}
\caption{Radius evolution of a $9.40\,\mathit M_{\odot}$ RSG during rapid envelope removal to mimic a CE in 1D. The envelope of the RSG is artificially ejected on a dynamical timescale until the remnant core mass reaches different values of 2.97, 2.85, 2.80 and 2.75\,$\mathit M_{\odot}$. The evolution of the corresponding surface radius is indicated by the solid lines in different colours as a function of time since the onset of the 1D-CE. The end points of the radius evolution are marked by circles with corresponding edge colours. For comparison, the black dot-dashed line indicates the RSG radius at the onset of the 1D-CE simulation. The Roche lobe radius of the RSG remnant after the CE with a BH and a NS companion is represented by the rectangles filled with transparent and dark colours, respectively. }
\label{fig:response of Giant during CE}
   \end{center}
\end{figure}

The same post-helium-burning RSG of 9.4\,$\mathit {M_\odot}$ as implemented in 3D CE simulations presented by \citet{2022A&AMoreno} and Vetter et al. (in prep), is considered to be the primary star in our effective ‘1D-CE’ models. We mimic the 3D CE phase in 1D by artificially stripping the envelope of the primary RSG as a single star on a dynamical timescale in the MESA code \citep[version r12778,][]{2011ApJSPaxton,2013ApJSPaxton,2015ApJSPaxton,2018ApJSPaxton,2019ApJSPaxton}, resulting in a remnant RSG core of $2.97\,\mathit {M_\odot}$, just like the 3D model. During the envelope removal, the remnant RSG cores expand first and then shrink significantly in our models (Fig.\,\ref{fig:response of Giant during CE}). This remnant RSG core of 2.97\,$\mathit M_{\odot}$ is then put into an orbit of a = 37\,$\mathit R_\odot$ with a $5.0\,\mathit M_\odot$ BH companion, as found in the original 3D CE simulation (Sect.\,\ref{Arepo_method}). We find that the radius of the remnant RSG core exceeds its Roche lobe in the post-3D CE binary orbit. The same overfilling behaviour occurs in the case of the $1.40\,\mathit {M_\odot}$ NS companion because of a much smaller Roche lobe radius compared to the case of the BH companion. In the 3D CE simulation, this core expansion could not be observed because the core is replaced by a point particle. This means that more material than that in the 3D CE models participates in the dynamic plunge-in phase, and may ultimately be ejected from the system. Consequently, the 3D CE simulations underestimate the ejecta mass and overestimate the orbital separation after the dynamic plunge-in phase. 

The true final RSG remnant masses remain unknown and we assume values of $2.75\,\mathit M_\odot$, $2.80\,\mathit M_\odot$, and $2.85\,\mathit M_\odot$ that all result in stars that do not immediately overfill their Roche lobes, as shown in Fig.\,\ref{fig:response of Giant during CE} (the mass of the RSG helium core is $2.74\,\mathit M_\odot$). The orbital separation and related Roche lobes of the post-CE binaries are determined by the traditional $\alpha$-formalism in which the CE ejection efficiencies obtained from the 3D CE simulations in Vetter et al. (in prep) are adopted: $\alpha_{\rm CE} = 0.57$ for the NS companion and $\alpha_{\rm CE} = 0.50$ for the BH companion. In the traditional $\alpha$-formalism, a fraction $ \alpha_{\rm{CE}}$ of the released orbital energy is used to unbind the envelope when the companion spirals into the RSG. The envelope can be ejected successfully if the orbital energy released equals the binding energy of the envelope $E_{\rm bind}$, that is to say

\begin{equation} \label{eqs.2}
E_{\rm bind} = \alpha_{\rm{CE}} \left (- \frac{GM_{\rm{rem}}M_2}{2a_{\rm f}} + \frac{GM_{\rm 1}M_2}{2a_{\rm i}} \right ).
\end{equation}
The right-hand side represents the released orbital energy, and the binding energy $E_{\rm bind}$ of the envelope can be expressed as 
\begin{equation} \label{eqs.3}
E_{\rm bind} =  -\int_{M_{\rm{rem}}}^{M_1} \frac{Gm}{r}\,{\mathrm{d}}m +\alpha_{\rm th}\int_{M_{\rm{rem}}}^{M_1}u\,{\mathrm{d}}m, 
\end{equation}
which is commonly re-defined by
\begin{equation} \label{eqs.4}
E_{\rm bind} \equiv -\frac{GM_1M_{\rm env}}{\lambda R_1},
\end{equation}
where the ejected envelope mass is $M_{\rm env} \equiv M_1 - M_{\rm rem}$. 
In the equations above (Eqs.~\eqref{eqs.2}, \eqref{eqs.3}, and \eqref{eqs.4}), G is the gravitational constant, $M_1$ and $M_2$ are the masses of the RSG and the compact companion, respectively. $M_{\rm rem}$ is the remnant RSG core mass after the successful CE ejection, which can be $2.75\,\mathit M_\odot$, $2.80\,\mathit M_\odot$, and $2.85\,\mathit M_\odot$ in our models. $a_{\rm i}$ and $a_{\rm f}$ are the orbital separations of the pre- and post-CE binaries. The binding energy is integrated from the boundary of the RSG remnant core to its surface. A fraction $\alpha_{\rm th}$ of the internal energy $u$ can contribute to the envelope binding energy. The envelope binding energy is re-defined in Eq.~\eqref{eqs.4} with the $\lambda$-parameter, often used in population synthesis models. Here, we set $\alpha_{\rm th} = 1$, and $\lambda = 1.81$ (Vetter et al. (in prep)). According to the traditional $\alpha$ formalism for the post-CE binary with a remnant RSG core of $2.75\,\mathit{M_\odot}$, $2.80\,\mathit{M_\odot}$, and $2.85\,\mathit{M_\odot}$, the orbital periods after the CE interaction for a BH and NS companion are given as $P_{\rm orb,i}$ in Table\,\ref{tab:pre-after-MT}.

\subsection{Post-CE binary MT}
\label{mass transfer_method}
To investigate the effect of additional MT on the evolution of the post-CE binaries described in Sect.\,\ref{post-CE_method}, RSG remnants orbiting a BH or NS point mass are simulated in MESA r12778 \citep{2011ApJSPaxton,2013ApJSPaxton,2015ApJSPaxton,2018ApJSPaxton,2019ApJSPaxton}.
A large nuclear network of 151 isotopes, “mesa151.net”, is employed in our models. This network includes a series of weak reactions (e.g. electron capture reactions), which is crucial in the evolution of low-mass He stars approaching core collapse \citep{2001ApJHeger,2013ApJJones,2015ApJWoosley,2017MNRASMoriya}. Several parameters governing the mixing process are set as follows. A convective mixing-length parameter of 1.82 and the Ledoux criterion with a semiconvection efficiency parameter of 0.1 are adopted. A substantial thermohaline coefficient of 666 is used to be self-consistent with the RSG model at the onset of the 3D CE simulations. The wind loss of massive stars is still an unclear process in the stellar evolution \citep{2022ARA&AVink}. For the mass loss of the RSG remnant during the binary interaction, MESA's ‘Dutch’ wind prescription \citep{2009A&AGlebbeek} with a scaling factor of 1 is utilised in our models. For cool stars with effective temperature below $10\,000\,\rm K$, we employ the wind mass loss scheme proposed by \citet{1988A&ASdeJager}. For hot stars with effective temperature exceeding $10\,000\,\rm K$, we apply the mass loss rate of \citet{2001A&AVink} for the phase with the surface hydrogen mass fraction greater than 0.4, and then switch to \citet{2000A&ANugis} if the surface hydrogen mass fraction is lower. To help models to converge, the wind mass loss is switched off at the final stage when the centre temperature exceeds $10^9\,\rm K$. Any wind mass loss during this final stage is negligible because of the relatively brief remaining time (${\sim}\, 10\,\rm yr$), and they hardly influence the binary evolution. Type\,2 opacities in MESA are used for a metallicity of 0.02.

Our binary models of RSG remnants and BH (or NS) companions are computed similarly to the models in \citet{2021ApJJiang}. To help with numerical convergence when using the large nuclear reaction network during the late nuclear burning stages of massive star evolution, we apply operator splitting after core C burning, i.e. solving the stellar structure equations and the nuclear reactions is no longer done simultaneously but in an alternating pattern \citep{2023ApJSJermyn}. The ‘Kolb’ \citep{1990A&AKolb} mass-transfer scheme is adopted and angular momentum loss in our models is attributed to mass loss and GW emission. We assume that at most 50\% of the mass transferred from the RSG remnant is accreted by the compact companion in our models, where the accretion is limited to the Eddington accretion rate \citep{2003MNRASPodsiadlowski}, which can be expressed by
\begin{equation}
\label{eqs.5}
\dot{M}^{\rm NS}_{\rm edd} = \frac{4\pi c R}{\kappa}
\end{equation}
and 
\begin{equation}
\label{eqs.6}
\dot{M}^{\rm BH}_{\rm edd} = \frac{4\pi G M_{\rm BH}}{\kappa c \eta},
\end{equation}
for the NS and BH companion, respectively. The mass that is lost from the system is assumed to carry the specific angular momentum of the orbit of the accretor. Here G is the gravitational constant, c is the speed of light and the electron scattering opacity with a hydrogen mass fraction of $X = 0.7$ is applied in our model, $\kappa = 0.2\,(1+X)\,\rm cm^2\,g^{-1}$. For a NS with a radius of $10\,\rm km$, the mass accretion rate is $1.8\times10^{-8}\,\mathrm {M_{\odot}\, yr^{-1}}$. For a non-rotating BH, we find a mass accretion rate of $1.9\times10^{-7}\,\mathrm {M_{\odot}\, yr^{-1}}$ with the assumption of $\eta = 0.07$ \citep{2003MNRASPodsiadlowski}. Our binary models are evolved until core Si depletion.

\begin{figure}
\begin{center}
\includegraphics[width=0.5\textwidth]{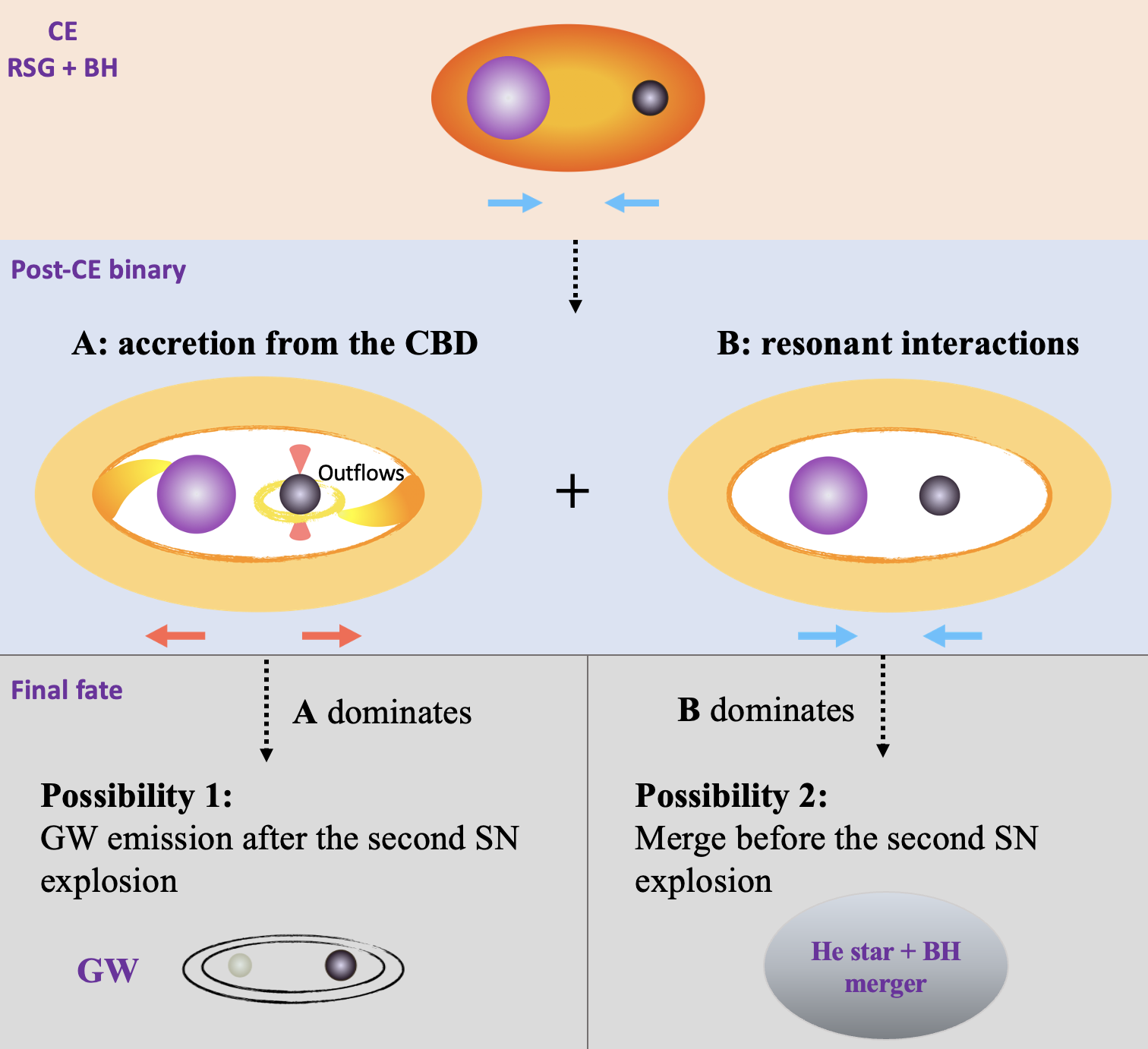}
\caption{Illustration of CBD-binary interactions in the post-CE binary with a BH companion. }
\label{fig:Disk_cartoon}
   \end{center}
\end{figure}

\subsection{CBD-binary interactions}
\label{CBD_methods}
Employing analytical models, we systematically explore the implications of CBD-binary interactions on the orbital evolution of the post-CE binaries without any binary MT. Using the total orbital angular momentum of a binary
\begin{equation} 
J_{\rm b} = \frac{M_{1}M_{2}}{M_{1} + M_{2}} \sqrt{G(M_{1} + M_{2})a},
\end{equation}
the orbital changes can be calculated from
\begin{equation} 
\label{eqs.8}
\frac{\dot{a}}{a} = \frac{\dot{M}_{\rm b}}{M_{\rm b}} + 2\frac{\dot{J}_{\rm b}}{J_{\rm b}} - 2\frac{\dot{M}_1}{M_1} - 2\frac{\dot{M}_2}{M_2}.
\end{equation}
Here, $M_{\rm 1}$, $M_{\rm 2}$ and $M_{\rm b}$ are the mass of the primary star, of the compact object and the total mass of the binary, respectively, and the corresponding mass changes over time are denoted by $\dot{M}_1$, $\dot{M}_2$, and $\dot{M}_{\rm b}$. $J_{\rm b}$ represents the orbital angular momentum that may change with time following $\dot{J}_{\rm b}$. The final separation $a$ can be obtained taking the separation changes over time $\dot{a}$ into account. Two aspects of interactions between the binary and CBD are discussed in this paper as illustrated in Fig.\,\ref{fig:Disk_cartoon}: (A) accretion from the CBD and (B) resonant interactions. All of these physical parameters in Eq.\,\eqref{eqs.8} can be modified by the CBD-binary interactions as elaborated in Sect.\,\ref{Accretion from disk_method} and Sect.\,\ref{Resonant_method}. 

In addition to the unclear mass accretion onto the binary because of the uncertain CBD properties (e.g. viscosity), jet-like bipolar outflows observed in the 3D CE simulations (\citet{2022A&AMoreno} and Vetter et al. (in prep)), and wind mass loss \citep{2023ApJTuna} can take away mass and angular momentum from the disk, giving rise to uncertainties of the total mass and lifetime of the CBD. According to \citet{2023ApJTuna}, the CBD lifetimes can be comparable to the thermal timescale of the remnant He star and even reach its nuclear timescale. Therefore, in our models, the CBD upper lifetime limit is $10^{5}\,\rm yr$, which is approximately the remaining time of the $2.8\mathit \,M_{\odot}$ RSG remnant until its core collapse ( ${\sim}\, 6\times10^{4}\,\rm yr$). To investigate the effect of CBD-binary interactions, we compute a grid of CBD models with the lifetimes $\tau_{\rm D}$ of $10^2,\ 10^3,\ 10^4,\ 5{\times}10^4$ and $10^5\,\rm yr$. The CBD mass $M_{\rm D}$ for the BH companion is $0.0001,\ 0.001,\ 0.01,\ 0.1,\ 0.2,\ 0.3,\ 0.4,\ 0.5,\ 0.6$ and $0.7\,\mathit M_{\odot}$. In the case of the NS companion, a narrower mass range is considered according to the outcomes of the 3D CE simulations, namely $0.0001,\ 0.001,\ 0.01,\ 0.05,\ 0.1,\ 0.15$ and $0.2\,\mathit M_{\odot}$. A constant average mass loss rate of the CBDs is assumed in our model, given by $\dot{M}_{\rm D} = M_{\rm D}/\tau_{\rm D}$.

\subsubsection{Accretion from the CBD (A)}
\label{Accretion from disk_method}

Angular momentum of gas in the CBD is transported to outer regions and gas losing angular momentum can thus migrate towards the inner edge of the CBD and potentially be accreted onto the binary \citep{2023ARA&ALai}. Recent hydrodynamical simulations of CBDs demonstrate that gas accretion can impart a positive torque onto the inner binary, leading to orbital widening of the inner binary \citep{2019ApJMunoz,2020ApJMunoz, 2023MNRASSiwekb}. To describe this process, we assume that material from the CBD is accreted by the inner binary with a rate of $\dot{M}_{\rm D}$, which is divided into two parts:
\begin{equation} 
\label{eqs.9}
\dot{M}_{\rm D} = \dot{M}_{\rm 1,b} +\dot{M}_{\rm 2,b}.
\end{equation}
Here, $\dot{M}_{\rm 1,b}$ and $\dot{M}_{\rm 2,b}$ are the fractions of material accreted onto the different components, which follow the preferential accretion from \citet{2023MNRASSiweka}: $\lambda_{\rm acc} = {\rm min}\left(\dot{M}_{\rm 2,b} / \dot{M}_{\rm 1,b}, \dot{M}_{\rm 1,b} / \dot{M}_{\rm 2,b} \right)$. This preferential accretion is a function of the mass ratio $q = {\rm min}\left(M_{1}/M_{2},M_{2}/M_{1}\right)$ and eccentricity of the binary. For circular binaries, the preferential accretion rate only depends on the mass ratio: $\lambda_{\rm acc} = q^{-0.9}$.

The accreted specific angular momentum onto the inner binary along the mass accretion is assumed to be the critical specific angular momentum $h_{\rm crit}$ (Eq.\,\eqref{eqs.1}), hence the angular momentum accretion rate is
\begin{equation} 
\label{eqs.10}
\dot{J}_{\rm acc} = h_{\rm crit}\,\dot{M}_{\rm D}.
\end{equation}

All of the material coming to the RSG remnant is assumed to be accreted: $\dot{M}_1 = \dot{M}_{\rm 1,b}$. Additionally, we assume mass loss with a rate of $\dot{M}_{\rm 2,loss}$ from the compact companion if the accretion onto it exceeds its Eddington limit,
\begin{equation} 
\label{eqs.11}
 \dot{M}_{\rm 2,loss} = -{\rm max}\left( \dot{M}_{\rm2,b} - \dot{M}_{\rm 2,edd},\,0 \right).
\end{equation}
Thus, only the mass accretion rate of $\dot{M}_{\rm 2}$ can be accreted by the compact companion as follows 
\begin{equation} 
\label{eqs.12}
\dot{M_2} = \dot{M}_{\rm2,b} + \dot{M}_{\rm 2,loss}.
\end{equation}

The angular momentum taken away by this mass loss is the specific orbital angular momentum of the compact companion in its relative orbit around the centre of mass as expressed by
\begin{equation} 
\label{eqs.13}
\dot{J}_{\rm 2,loss} = \left(\frac{a M_1}{M_1 + M_2} \right)^2 \Omega_{\rm b} \dot{M}_{\rm 2,loss}.
\end{equation}
Here, $\Omega_{\rm b}$ is the orbital angular velocity of the inner binary. Only considering mass accretion from the CBD, the change of the orbital angular momentum in Eq.\,\ref{eqs.8} can be expressed as $\dot{J}_{\rm b} =\dot{J}_{\rm acc} +\dot{J}_{\rm 2,loss}$.

\subsubsection{Resonant interactions (B)}
\label{Resonant_method}

Orbital angular momentum of the binary can be exchanged with particles in the CBD because of gravitational tidal torques between the disk and the binary, which can be strongest for certain resonances \citep{1996ASICLubow,2000prplLubow}. As discussed in Sect.\,\ref{Accretion from disk_method}, angular momentum in the disk is transported outwards via viscous torques. An equilibrium state at the inner edge of the disk is reached when the gravitational tidal torque is comparable to the viscous torque \citep{2023ARA&ALai}.

The change in the binary orbit due to the resonant interactions between the binary and the CBD \citep{1996ASICLubow,2000prplLubow,2001ApJSpruit} can be estimated by
\begin{equation} 
\label{eqs.14}
\frac{\dot{a}_{\rm CB}}{a} = 2 \frac{\dot{J}_{\rm CB}}{J_{\rm b}} = -\,6 \frac{l}{m} \frac{\alpha}{\sqrt{1-e^2}\,\mu_1\,(1-\mu_1)}\left( \frac{H_{\rm in}}{R_{\rm in}}\right)^2 \frac{a}{R_{in}}\,q_{\rm d}\,\Omega_{\rm b}.
\end{equation}
Here, the mass ratio of the CBD and inner binary is $q_{\rm d} = \pi {R_{\rm in}}^2 \sigma / M_{\rm b}$, where $\sigma$ describes the disk surface density profile. The binary mass parameter $\mu_1$ is defined as $\mu_1 = M_2/M_{\rm b}$, the viscosity parameter is assumed as $\alpha = 0.1$ in our models, and the aspect ratio of the disk is $H/R = H_{\rm in}/R_{\rm in}= 0.1$, where R and $R_{\rm in}$ are the half angular-momentum radius and the inner radius of the disk, respectively. $e$, $a$, and $\Omega_{\rm b}$ represent the binary eccentricity, separation and orbital angular velocity, respectively. The azimuthal harmonic number and time-harmonic number are represented by $m$ and $l$, respectively, if the gravitational potential produced by the binary is expended in a Fourier series \citep{1980ApJGoldreich,2000prplLubow}. According to the SPH simulations studied by \citet{1991ApJArtymowicz}, the outer Lindblad resonance of $l = 1$ and $m = 2$ is dominant, driving the equilibrium at the inner edge of the CBD \citep{2000prplLubow}. 
Moreover, the resonance can pump eccentricity \citep{1996ASICLubow,2000prplLubow,2013A&ADermine} with a rate of 
\begin{equation} 
\label{eqs.15}
{ \dot{e} = 
    \left\{ 
    \begin{matrix}      
    2\frac{(1-e^2)}{e + \frac{\alpha}{100e}}\left ( \frac{l}{m} - \frac{1}{\sqrt{1-e^2}} \right )\frac{\dot{a}_{\rm CB}}{a} , &  e \leq 0.2 \\
    (\frac{C_{1}}{e} + C_{2})\frac{\dot{a}_{\rm CB}}{a}, & 0.2 < e \leq 0.7 \\
    0, & e > 0.7
    \end{matrix}
    \right.}
\end{equation}
where $C_1 = -1.3652$ and $C_2 = 1.9504$ are constants. For small or moderate eccentricities ($e < 0.2$), the eccentricity increases with a high pumping efficiency due to the strong effect of the $l = 1$ and $m = 2$ resonance. However, as the eccentricity increases, resonances that damp the eccentricity become more and more important, and the eccentricity can eventually reach an equilibrium value of $~0.5-0.7$ \citep{2000prplLubow}, in line with the outcome of 2D CBD simulations \citep{2023MNRASSiwekb}. The rate of change in eccentricity thus drops as a function of $1/e$ for larger eccentricities ($e\,{\ge}\,0.2$), and the eccentricity reaches an equilibrium at a maximum value of 0.7 in this model.

We match the total mass and angular momentum of a Keplerian CBD to those of the bound material at the end of the 3D CE simulations ($M_{\rm D}$ and $J_{\rm D}$, see Sect.\,\ref{Arepo_method}), 
\begin{equation} 
\label{eqs.16}
M_{\rm D} = \int_{\rm R_{in}}^{\rm R_{out}} \sigma\,2\pi r\, {\mathrm d}r
\end{equation}
and
\begin{equation} 
\label{eqs.17}
J_{\rm D} = \int_{\rm R_{in}}^{\rm R_{out}} r \sqrt{\mathit{\frac{\rm G \left( M_1 + M_2\right)}{r}}} \, \sigma\,2\pi r\, {\mathrm d}r,
\end{equation}
where ${\rm R_{in}}$ and ${\rm R_{out}}$ are the inner and outer boundaries of the CBD. The inner radius of the disk changes with the eccentricity, separation of the binary and disk structure \citep{1994ApJArtymowicz, 2020A&AOomen}, which can be expressed as

\begin{equation} 
\label{eqs.18}
R_{\rm in} = a \left[ 1.7 + \frac{3}{8} \sqrt{e}\,{\rm log} \left(\alpha^{-1} \left( \frac{H}{R} \right)^{-2}  \right)\right].
\end{equation}

The surface density distribution $\sigma$ in the CBD is assumed to follow a power-law distribution \citep{2015A&AVos,2020A&AOomen} of the form
\begin{equation}
\label{eqs.19}
\sigma(r) = \frac{D_c}{r^\delta},
\end{equation}
where we set $\delta = 1.5$, and $D_{\rm c}$ and $R_{\rm out}$ can be constrained by the time-dependent angular momentum and mass in the CBD from Eqs.~\eqref{eqs.16} and \eqref{eqs.17}.

In summary, for the model of resonant interactions, the total mass and angular momentum of the CBD in our model decrease with time following the rate of $\dot{M}_{\rm D}$ and $\dot{J}_{\rm acc}$, which can influence the surface density distribution ($\sigma$) of the CBD (see Eqs.~\eqref{eqs.16}, \eqref{eqs.17}, \eqref{eqs.18} and \eqref{eqs.19}). As a function of $\sigma$, $q_{\rm d}$ in Eq.\,\eqref{eqs.14} is therefore time-dependent. Furthermore, eccentricity and $R_{\rm in}$ change with time following Eqs.\,\eqref{eqs.15} and \eqref{eqs.18}. The mass of each component in the binary remains constant if only the effect of the resonant interactions is considered, and $\dot{J}_{\rm CB}$ is the only component contributing to the changes of angular momentum of $\dot{J}_{\rm b}$ in Eq.\,\eqref{eqs.8}.

Overall CBD-binary effects include the mass accretion from the CBD (A) and the resonant interactions (B). In this case, the $\dot{J}_{\rm b}$ in Eq.\,\eqref{eqs.8} should be $\dot{J}_{\rm b} = \dot{J}_{\rm acc} + \dot{J}_{\rm 2,loss} + \dot{J}_{\rm CB}$, and $\mu_{\rm 1}$ in Eq.\,\eqref{eqs.14} is time-dependent because of the mass accretion from the CBD. In Sect.\,\ref{CBD-binary interactions}, we demonstrate how the mass accretion from the CBD and the resonant interactions influence the evolution and final fate of the post-CE binary without any binary interactions, respectively. In addition, the overall effect including both is addressed.

\section{Additional binary MT}
\label{Additional mass transfer}

\subsection{Two fiducial models}
\label{Two fiducial models}

\begin{figure}
\begin{center}
\includegraphics[width=0.5\textwidth]{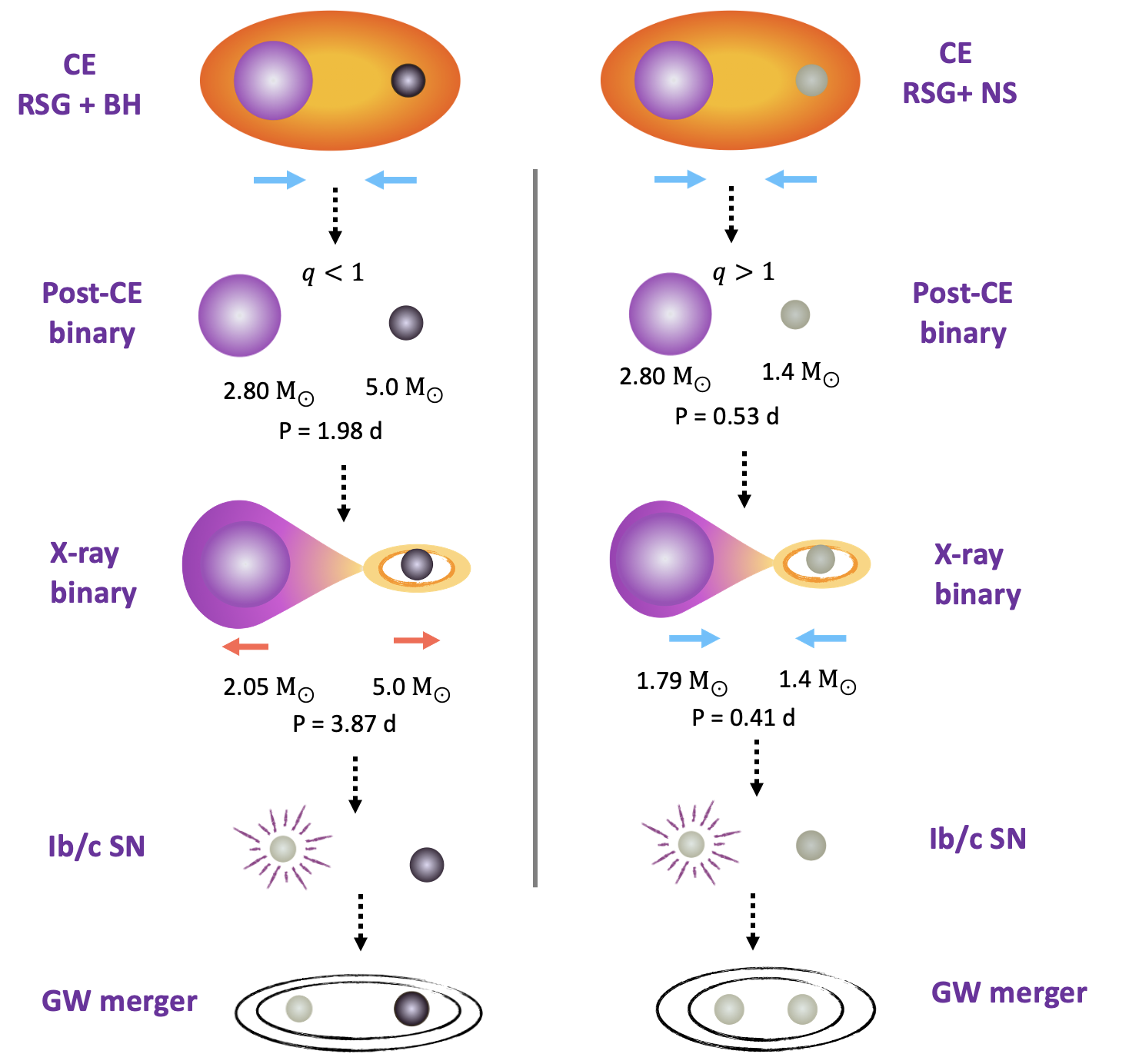}
\caption{Illustration of the evolution of post-CE binaries undergoing the additional MT, ultimately leading to the formation of double compact objects and producing GW emission.  }
\label{fig:MT_cartoon}
   \end{center}
\end{figure}

The remnant RSG core of $2.8\,\mathit {M_\odot}$ is considered as our fiducial model, interacting with a $5.0\,\mathit M_{\odot}$ BH and $1.4\,\mathit M_{\odot}$ NS companion in the subsequent evolution. The initial orbital periods of these post-CE binaries are $1.98\,\rm d$ and $0.53\,\rm d$, respectively. In this section, we will describe how the additional binary MT influences the evolution and final fate of these systems, which is illustrated in Fig.\,\ref{fig:MT_cartoon}.

\begin{figure*}
\begin{center}
\includegraphics[width=1.0\textwidth]{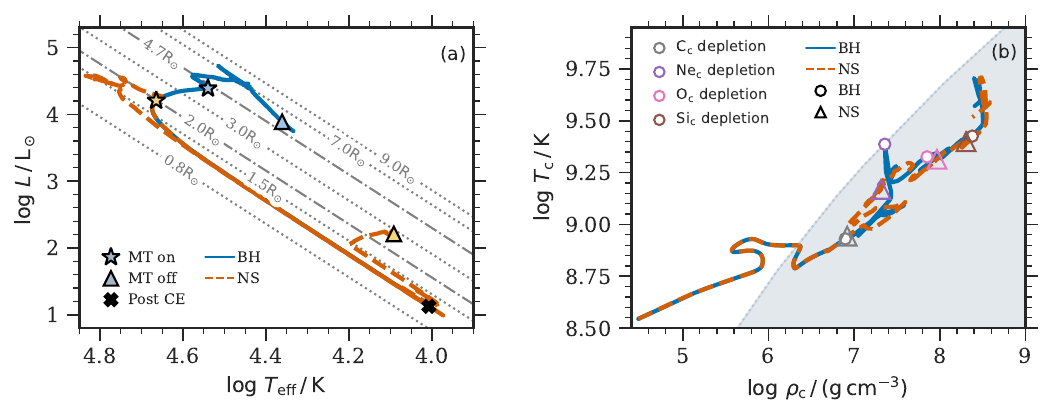}
\caption{Evolutionary tracks of RSG remnants with a BH and NS companion in the Hertzsprung-Russell diagram (panel a) and in the plane of central temperature versus central density (panel b). The solid blue and dashed orange lines in panel a indicate the evolutionary tracks of RSG remnants undergoing MT to a BH and NS companion, respectively. The black cross represents the RSG remnant at the end of the CE phase, namely the start of evolutionary tracks. The onset and termination of MT are marked with symbols of stars and triangles, respectively. Constant radii are shown by grey lines, among which the dash-dotted lines are the Roche lobe radii given the initial configuration of the post-CE binaries. In panel b, different colours mark the evolutionary points at which the corresponding abundance depletes in the centre of stars. The grey zone indicates the centre of the core is in an electron degenerate state.}
\label{fig:HR_T_rho}
   \end{center}
\end{figure*}

Fig.\,{\ref{fig:HR_T_rho} shows the evolutionary tracks of the RSG remnants in the Hertzsprung-Russel (HR) diagram and the central temperature-density ($T_{\rm c}-\rho_{\rm c}$) plane. After an initial thermal adjustment phase, characterised by a rapid increase in luminosity at constant radii, the RSG remnants expand, as shown in Fig.\,{\ref{fig:HR_T_rho}{\color{blue}{a}}. As a result, the RSG remnants overfill their Roche lobe radii within ${\sim}\, 5000\rm\,yr$ for the BH companion and ${\sim}\, 20\rm\,yr$ for the NS companion. The post-CE binary with a NS companion initiates the thermal-timescale MT earlier than with a BH companion because of a smaller Roche lobe radius and tighter orbit after the CE phase. The RSG remnant with a NS companion becomes hotter compared to that with a BH companion because more of the envelope is stripped during the MT. In the late evolution, the energy in the RSG remnants is used to expand the star, hence the total luminosities decrease in both cases. The RSG remnants evolve to Si depletion in both cases as illustrated in Fig.\,{\ref{fig:HR_T_rho}{\color{blue}{b}}. The evolutionary tracks overlap in both cases before C depletion, while variations occur during the subsequent stages of evolution because of the different MT processes.

\begin{figure*}
\begin{center}
\includegraphics[width=1.0\textwidth]{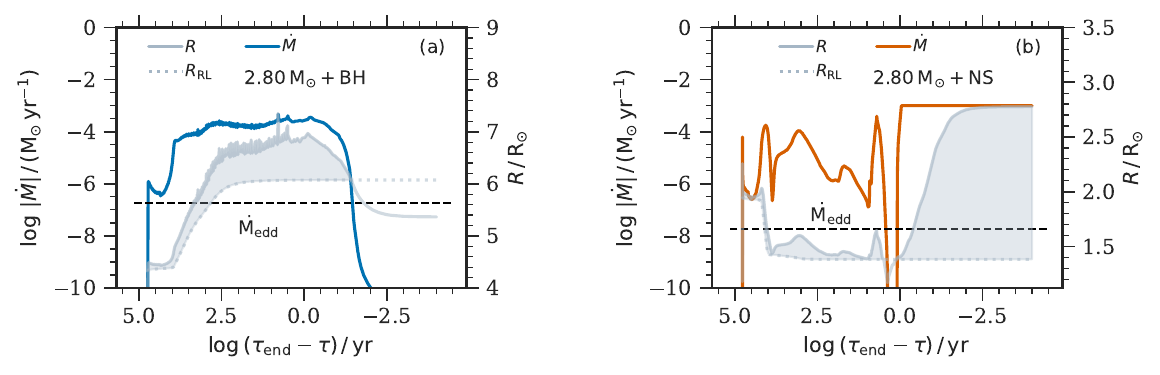}
\caption{Mass-transfer rate and (Roche lobe) radii of the RSG remnants orbiting around a BH (panel a) and a NS companion (panel b) as a function of the remaining time until the end of the computation. The solid and dotted grey lines denote radii of RSG remnants and Roche lobe radii, respectively. Mass-transfer rates of the post-CE binary with a BH and a NS companion are shown by the blue and orange lines, respectively. The horizontal dashed black lines indicate the Eddington-limited accretion for different compact objects ($1.9 \times 10^{-7}\,\mathrm{ M_{\odot}\, yr^{-1}}$ in the panel a for the BH; $1.8 \times 10^{-8}\,\mathrm{  M_{\odot}\, yr^{-1}}$ in the panel b for the NS). The shaded grey areas illustrate how much the radii of RSG remnants exceed their Roche radii.}
\label{fig:mass_loss_t_and_R}
   \end{center}
\end{figure*}

Binary MT rates are given in Fig.\,{\ref{fig:mass_loss_t_and_R}}, and are determined by the difference of the surface and the Roche lobe radius. The Roche lobe radius changes over time during the initial stages of MT, while it remains relatively stable once the mass ratio of the binary reaches an almost constant value. Therefore, the variations of surface radius drive the variations in MT rate in the later phase. A similar phenomenon is observed in the post-CE binary with a NS companion. The radial contraction and expansion are induced by the different time-dependent burning zones inside the core and can be understood from the mirror principle. More details can be found in Wei et al. (in prep) \citep[see also][]{2013ApJTauris}. In the post-CE binary with a NS companion, two distinct phases of MT take place. The first MT phase terminates at ${\sim}\, 2\,\rm yr$ prior to core collapse, while the second MT phase occurs shortly before core collapse as a result of re-expansion.

During the MT phase, mass accretion onto the compact companion is restricted by the Eddington-limited accretion rates specified in our models, which are $1.8\times10^{-8} \, \mathrm {M_\odot \,  yr^{-1}}$ and $1.9\times10^{-7}\, \mathrm{M_\odot \, yr^{-1}}$ in the case of the NS and BH companion, respectively. The MT rate during the binary evolution is higher by nearly $2-4$ orders of magnitude than the corresponding Eddington limits indicated by black dashed lines in Fig.\,\ref{fig:mass_loss_t_and_R}. This means that most of the transferred material exceeds the Eddington limit and is re-emitted by the compact companion. In fact, only ${\sim}\, 0.001\,\mathit M_\odot$ or ${\sim}\, 0.01\,\mathit M_\odot$ is accreted by the compact objects in our models (Table\,\ref{tab:pre-after-MT}).

\begin{figure*}
\begin{center}
\includegraphics[width=1.0\textwidth]{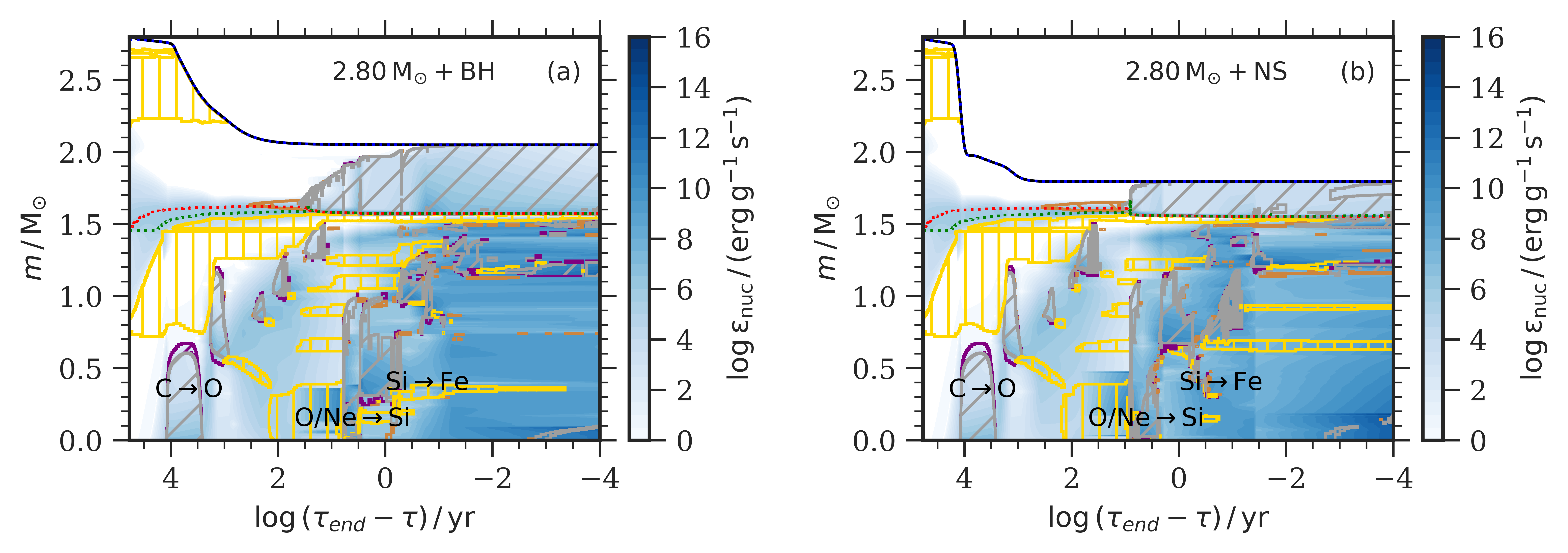}
\caption{Kippenhahn diagram of the RSG remnant of $2.80\,\mathit M_{\odot}$ with a BH and NS companion after the CE phase. The evolution of the interior of the RSG remnants through MT to a BH and NS companion are shown in the panel a and panel b, respectively, as a function of the remaining time until central Si depletion (Si mass fraction $< 10^{-3}$). The blue colours indicate the intensity of nuclear energy production. Convection, convective overshooting, semi-convection, and thermohaline mixing are marked with grey, purple, brown, and yellow hatched zones, respectively. The red and green dotted lines denote the C and O core boundaries\protect\footnotemark[1].}
\label{fig:kipps}
   \end{center}
\end{figure*}

\footnotetext[1]{Boundary definition: the outermost location in mass coordinates where the mass fraction of the corresponding abundance (e.g. $^{12}\textrm{C}$ abundance for C core) exceeds 0.1, and meanwhile, that of the prior depleted abundance (e.g. $^{4}\textrm{He}$ abundance for C core) decreases to less than 0.5 firstly.}

The MT episodes in the two post-CE binaries lead to different final binary configurations (see the illustration in Fig.\,\ref{fig:MT_cartoon}). The RSG remnant of $2.80\,\mathit M_{\odot}$ after the CE simulation has a slight hydrogen-rich layer of ${\sim}\, 0.06\,\mathit M_{\odot}$ above the helium-rich envelope and a CO core of $1.45\,\mathit M_{\odot}$. During core-C burning, the CO core grows because of He-shell burning, leading to almost the same final CO core mass of $1.56\,\mathit M_{\odot}$ at the end of the computations in both models. As shown in Fig.\,\ref{fig:kipps}, the ${\sim}\, 0.23\,\mathit M_{\odot}$ remaining envelope of the RSG remnant with a NS companion is less than the ${\sim}\, 0.48\,\mathit M_{\odot}$ remaining envelope with a BH companion. This discrepancy arises from a more significant material transfer towards the NS companion during RLOF (Fig.\,\ref{fig:mass_loss_t_and_R}). Consequently, the final mass of the RSG remnants with a BH and NS companion are $2.05\,\mathit M_{\odot}$ and $1.79\,\mathit M_{\odot}$, respectively. The binary MT shrinks the initial orbit to a period of $0.41 \rm\,d$ in the case of the NS companion, while the orbit widens by a factor of 2 to a period of $3.87 \rm\,d$ for the BH companion. The distinct behaviours of separation in different binary configurations can be explained by the angular momentum exchange. The specific angular momentum around the more massive component in its relative orbit around the centre of mass is smaller than that of the less massive one. Therefore, in the case of fully conservative MT, the separation of the binary tends to shrink to keep the total angular momentum unchanged when the material is transferred to the less massive component from the more massive star, and the orbit expands when the masses of the two components reverse. For non-conservative MT, such as our models, a similar tendency is observed. However, the transition mass ratio is not 1.0 because of the mass lost from the system (see Fig.\,\ref{fig:a_M1_beta}).

The MT alters the amount of envelope mass and hence gives rise to subtle changes in the internal stellar structure (Fig.\,\ref{fig:kipps}). With an initial CO core of around $1.45\,\mathit  M_{\odot}$, the central temperature in both cases ($\rm log\,\mathit T_c > 8.75\ K$) is sufficiently high to initiate carbon burning in the core (see Fig.\,\ref{fig:kipps} and Fig.\,\ref{fig:HR_T_rho}{\color{blue}{b}}). Subsequently, different C-shell burning advances outwards during the evolution. The RSG remnants in both cases finish most of their life with a degenerate core after core-C burning (Fig.\,\ref{fig:HR_T_rho}{\color{blue}{b}}). The occurrence of off-centre Ne ignition in these two cases is attributed to the temperature inversion induced by neutrino losses in the core. Ne ignition is close to the centre of the core, and the burning front then moves in and outwards. Other advanced burning stages (i.e. O, Si) ignite in the core subsequently. During the Ne/O burning, convective zones extend from the He-burning shell to the whole envelope. Meanwhile, the surface radius expands (see Fig.\,\ref{fig:mass_loss_t_and_R}) because of the feedback from He-shell burning.

The final ONeMg cores in our models have a mass exceeding $1.37 \,\mathit M_\odot$, and they undergo the O/Ne/Si burning process in their cores, ultimately leading to the formation of iron cores. Therefore, our models are predicted to undergo iron core-collapse supernovae (Fe CCSNe) rather than electron capture supernovae (ECSNe) \citep{1987ApJNomoto, 2004ApJPodsiadlowski,2013ApJTakahashi,2015MNRASTauris}. In our models, the remaining He-rich envelope masses are ${\sim}\, 0.23\,\mathit M_\odot$ and ${\sim}\, 0.48\,\mathit M_\odot$ in the case of NS and BH companion, respectively. They are more likely to explode as Type Ib/c SNe, rather than so-called ultra-stripped SNe, for which the remaining He-rich envelope mass is defined as less than ${\sim}\, 0.20\,\mathit M_\odot$ in \citet{2013ApJTauris,2015MNRASTauris}.

\subsection{Models with different remnant masses}
\label{Models with different remnant masses}

We use different RSG remnants of $2.75\, \mathit M_\odot$, $2.80\, \mathit M_\odot$ and $2.85\, \mathit M_\odot$ after the CE phase because of the uncertainties in the 3D CE simulations (Sect.\,\ref{post-CE_method}). Employing the traditional $\alpha$-formalism, different post-CE binary configurations are found (Sect.\,\ref{post-CE_method}), which give rise to different MT processes. The initial and final properties of all post-CE binaries prior to core collapse are summarised in Table\,\ref{tab:pre-after-MT}.

\begin{table*}[h]
    \renewcommand  
    \arraystretch{1.5} 
    \centering
    \setlength\tabcolsep{5.0pt}
    \caption{\label{tab:pre-after-MT}Properties of post-CE binaries just after the CE phase (marked by ‘i’), and prior to core collapse (indicated by ‘f’).}
    \begin{tabular}{c c c c c c c c c c c c c c}
    \hline 
    \hline 
     Binary systems            & $M_{\rm 1,i}$     & $P_{\rm orb,i}$ &  $M_{\rm core,i}^{\rm CO}$ & $M_{\rm 1,f}$     & $P_{\rm orb,f}$  &  $M_{\rm core,f}^{\rm CO}$  & $M_{\rm He,f}^{\rm env}$  & $\Delta M_{\rm NS\ or\ BH}$ & SNe type &  $\tau_{\rm grw}$ & $f_{\rm merger}$ & $f_{\rm bound}$ &$f_{\rm disrupt}$ \\ 
     & $(\rm M_{\odot})$      & $(\rm days)$    &  $(\rm M_{\odot})$ &   $(\rm M_{\odot})$  & $(\rm days)$     &  $(\rm M_{\odot})$ &   $(\rm M_{\odot})$ & $(\rm M_{\odot})$ &   & $\rm Gyr$ & $\%$ & $\%$ & $\%$\\ 
    \hline  
    2.85\,$\rm M_{\odot}$ + BH & 2.85  & 4.39   & 1.45  & 2.48  & 5.76  & 1.57  & 0.91 & 1.04e-2 & Ib/c            & 2952 & 5  & 21 & 79 \\
    
    2.85\,$\rm M_{\odot}$ + NS & 2.85  & 1.25   & 1.45  & 1.85  & 0.94  & 1.56  & 0.29 & 1.00e-3 & Ib/c            & 61    & 20 & 35 & 65 \\  
    
    2.80\,$\rm M_{\odot}$ + BH & 2.80  & 1.98   & 1.45  & 2.05  & 3.87  & 1.56  & 0.48 & 9.66e-3 & Ib/c            & 763  & 9  & 27 & 73\\ 
    
    2.80\,$\rm M_{\odot}$ + NS & 2.80  & 0.53   & 1.45  & 1.79  & 0.41  & 1.56  & 0.23 & 1.00e-3 & Ib/c            & 6     & 44 & 48 & 52\\
    
    2.75\,$\rm M_{\odot}$ + BH & 2.75  & 0.74   & 1.45  & 1.88  & 1.69  & 1.56  & 0.31 & 2.68e-3 & Ib/c            & 74   & 21 & 41 & 59\\  
    
    2.75\,$\rm M_{\odot}$ + NS & 2.75  & 0.19   & 1.45  & 1.71  & 0.15  & 1.54  & 0.16 & 4.96e-4 & ultra-stripped  & 0.37  & 64 & 64 & 36 \\     
    \hline 
    \end{tabular}
    
{\textbf{Notes.}     
         \footnotesize               
             Initial properties of post-CE binary: $M_{\rm 1,i}$ -- initial mass of the RSG remnant, $P_{\rm orb,i}$ -- initial orbital period, $M_{\rm core,i}^{\rm CO}$ -- initial CO core mass of the RSG remnant.
             Final properties prior to core collapse: $M_{\rm 1,f}$ -- final mass of the RSG remnant, $P_{\rm orb,f}$ -- final orbital period, $M_{\rm core,f}^{\rm CO}$ -- final CO core mass of the RSG remnant, $M_{\rm He,f}^{\rm env}$ -- final mass of the He-rich envelope, $\Delta M_{\rm NS\ or\ BH}$ -- the total amount mass accreted by the compact companion, SNe type -- the SNe type when the star explodes, $\tau_{\rm grw}$ -- GW merger timescale of final double compact objects without NS kicks, $f_{\rm merger}$ -- GW-induced merger fraction of final double compact objects within a Hubble time considering NS kicks, $f_{\rm bound}$ -- fraction of bound binary systems after NS kicks, $f_{\rm disrupt}$ -- fraction of disrupted binary systems after NS kicks
	}
\end{table*}

\begin{figure}
\begin{center}
\includegraphics[width=0.5\textwidth]{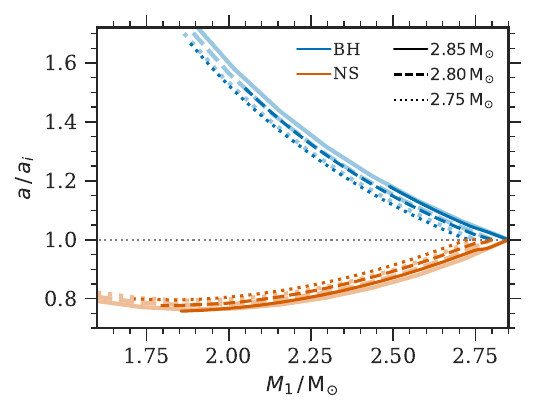}
\caption{Evolution of the ratio of present-day to initial orbital separation of post-CE binaries undergoing MT from different primaries to a BH or NS companion as a function of the primary mass $M_1$. The solid, dashed, and dotted lines denote the different primary stars. Binaries with a BH and a NS companion are represented by the blue and orange lines, respectively. For comparison, we show the analytical non-conservative orbital evolution according to \citet{1996A&ATauris} with transparent lines, and $\beta$ is assumed to be 0.85. }
\label{fig:a_M1_beta}
   \end{center}
\end{figure}

The orbital evolution, quantified by the ratio of the orbital separation and its initial value, is shown as a function of the time-dependent RSG remnant mass in Fig.\,\ref{fig:a_M1_beta}. For the same compact companion (a BH or NS), a less massive RSG remnant (e.g. $2.75\, \mathit M_\odot$) transfers more mass to its companion because of the corresponding tighter orbit at the onset of MT (see Table\,\ref{tab:pre-after-MT}). For the same reason, the post-CE binary with a NS companion undergoes more envelope stripping during the binary interaction than for a BH companion, given the fixed RSG remnant.

The separation of the post-CE binary with a BH companion increases during the binary MT phase. The orbital expansion due to the MT phase is much more efficient in the post-CE binary with a less massive RSG remnant, because of the larger ratio of the initial mass ratio to its final value (cf. Eq.\,\eqref{eqs.20}). For example, the orbit of the post-CE binary widens by ${\sim}\, 67\%$ in the case of a $2.75\, \mathit M_\odot$ RSG remnant, while it only widens by ${\sim}\, 18\%$ for the $2.85\, \mathit M_\odot$ + BH binary. For the post-CE binary with a NS companion, the orbital shrinking slows down and then goes up slightly, as the RSG remnant loses mass. In the case of $2.85\, \mathit M_\odot$ RSG remnant, the orbit eventually shrinks by ${\sim}\, 25\%$. 

For comparison, we show the orbital evolution induced by non-conservative interactions following the analytical method in \citet{1996A&ATauris} given by
\begin{equation} 
\label{eqs.20}
 \frac{a_{\rm f}}{a_{\rm i}} = \left ( \frac{q_{\rm i}\,(1-\beta)+1}{q_{\rm f}\,(1-\beta)+1}\right )^{\frac{3\beta-5}{1-\beta}}\left( \frac{q_{\rm i}+1} {q_{\rm f}+1}\right ) \left(\frac{q_{\rm i}}{q_{\rm f}} \right )^2,
\end{equation}
where the ratio of orbital separation with respect to its initial value $a_{\rm f}/a_{\rm i}$ is a function of mass ratio $q = M_{\rm donor}/M_{\rm accretor} = M_1/M_2$. $q_{\rm i}$ and $q_{\rm f}$ are the initial and final mass ratios. $\beta $ represents the fraction of transferred material which is ejected isotropically around the compact companion. As shown in Fig.\,\ref{fig:a_M1_beta}, the transparent curves from the analytical method with $\beta = 0.85$ are in line with our outcome from the simulations.

\subsection{GW mergers in a Hubble time}
\label{GW merger rate for MT}

A supernova explosion has the potential to perturb the binary orbit, consequently influencing the likelihood of a subsequent GW merger event within a Hubble time. The complex explosion process is beyond the scope of this work. To simplify, we consider the impact of the mass loss and NS kick imparted by a SN explosion to figure out whether an ensuing double compact merger event driven by GW emission can occur within a Hubble time.

Disregarding the effects of NS kicks, the orbit can become wider and more eccentric because of the mass loss during an SN explosion alone. The final separation relative to its initial value can be expressed by
\begin{equation} 
\label{eqs.21}
\frac{a_f}{a_i} = \frac{M_1 + M_2 -\Delta M }{M_1 + M_2 -2 \Delta M},
\end{equation}
and the orbital eccentricity after the SN event is determined by
\begin{equation} 
\label{eqs.22}
e = \frac{\Delta M}{M_1+M_2-\Delta M}.
\end{equation}
Here, we assume that mass loss ($\Delta M$) from the exploding star with an initial mass of $M_1$ takes place instantaneously, perturbing the initial circular binary comprising a companion of mass $M_2$.

As an example, we take the fiducial model of $2.80\,\mathit M_{\odot}$ to investigate the effect of the mass loss during a SN explosion on the post-CE-mass-transfer (post-CE-MT) binary configuration, and on the subsequent GW merger events. After the additional MT discussed in Sect.\,\ref{Two fiducial models}, the final masses of RSG remnants with a BH and NS companion are $2.05\,\mathit M_{\odot}$ and $1.79\,\mathit M_{\odot}$, respectively. A NS of $1.40\,\mathit M_{\odot}$ is assumed to be formed following a SN explosion in both post-CE-MT binaries. In the case of a NS companion, the mass loss during the SN explosion is nearly $0.39\,\mathit M_{\odot}$, leading to a wider separation (increasing by a factor of 1.16) and an eccentricity of 0.13. The GW merger time of double compact objects left from the SN explosion can be estimated according to Eq.\,(5.14) in \citet{1964PhRvPeters}, which is about $6\,\rm Gyr$ in this case. Thus, the post-CE binary of $2.80\,\mathit M_{\odot}$ + NS, that experienced MT, could lead to a NS-NS merger within a Hubble time (${\sim}\, 13.7\,\rm Gyr$) even without kicks. However, in the case of the BH companion, the merger time of around $763\,\rm Gyr$ is too long to produce a NS-BH merger event within a Hubble time. In all of the models we computed (Table\,\ref{tab:pre-after-MT}), only two cases with NS companions can merge within a Hubble time without kicks. In the case of NS companion, the post-CE orbits are already tighter than for BH companion and the additional MT phase shrinks the binary orbit even further.

\begin{figure}
\begin{center}
\includegraphics[width=0.5\textwidth]{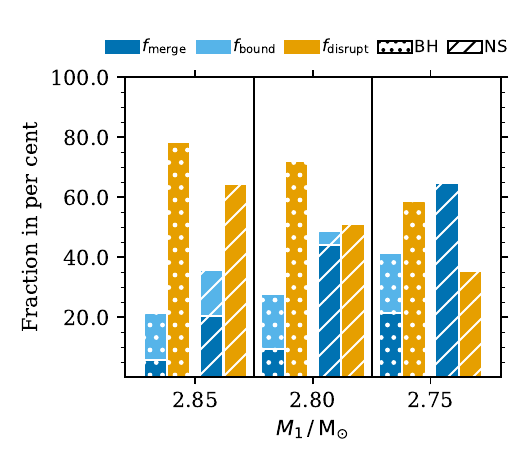}
\caption{Fraction of mergers within a Hubble time, bound, and disrupted systems for different post-CE binaries experiencing MT and NS kicks. }
\label{fig:sn-kicks-gw-merger}
   \end{center}
\end{figure}

A newly formed NS can receive a kick from a SN explosion, which may destroy the binary system or help in promoting double compact merger events. The discrepant influences depend on the magnitude and direction of the NS kick. To obtain the binary configurations (i.e. orbital separation and eccentricity) after a SN explosion taking both kicks and mass loss into account, we utilise the formalism developed by \citet{2002ApJPfahl}. We consider isotropic NS kicks, and a Maxwellian distribution with $\sigma_{\rm kick} = 265\,\rm km\,s^{-1}$ for the kick velocities, following the proper motion observations of pulsars in \citet{2005MNRASHobbs}. The effect of the SN blast on the compact companion is neglected \citep{2018ApJHirai}. With the configuration of the non-disrupted post-SN binary, GW merger times are computed following \citet{1964PhRvPeters}. Compared to the double compact merger fraction calculated by the formalism in \citet{1995MNRASBrandt}, nearly identical results are obtained with deviations of $0.01\%$ at most. However, we can take advantage of the formalism in \citet{2002ApJPfahl} to calculate the SN impacts on the eccentric binary for our eccentricity-pumping CBD models in the next section. 

The post-SN binaries can be disrupted by the SN explosion, survive as bound binary systems, and merge via GW emission within a Hubble time. Computed fractions of these outcomes are shown in Fig.\,\ref{fig:sn-kicks-gw-merger}. As we discussed above, post-CE binaries with BH companions undergoing additional MT cannot merge within a Hubble time without the influence of kicks. Taking into account the NS kicks makes it possible for these binaries to achieve mergers via GW emission within a Hubble time. Under the impact of NS kicks, $20\% - 40\%$ of the binary systems with BH companions remain bound, and $5.7\%$, $9.5\%$, and $21.5\%$ of $2.85\,\mathit M_{\odot}$ + BH, $2.80\,\mathit M_{\odot}$ + BH and $2.75\,\mathit M_{\odot}$ + BH binaries, respectively, merge within a Hubble time (see Fig.\,\ref{fig:sn-kicks-gw-merger}). The trend of an increasing merger fraction with decreasing donor mass is consistent with the corresponding orbital configurations. The binary systems with a less massive donor have tighter orbits before the SN explosion (Sect.\,\ref{Models with different remnant masses}), inducing a higher GW merger fraction. In the case of NS companions, the separation of the binary of $2.85\,\mathit M_{\odot}$ + NS is too wide to merge within a Hubble time without kicks, while about $20\%$ of such binary systems merge due to GW emission under the assumption of kicks. 

Kicks may help to facilitate double compact mergers as shown above; however, they can also work in the opposite direction. The binary systems of $2.80\,\mathit M_{\odot}$ + NS and $2.75\,\mathit M_{\odot}$ + NS are expected to merge within a Hubble time even without kicks. However, around $35\% - 50\%$ of these binary systems are disrupted by NS kicks, which significantly reduces their merger fraction (see Fig.\,\ref{fig:sn-kicks-gw-merger}). In the case of $2.75\rm\,M_{\odot}$ + NS, all of the surviving binaries become mergers within a Hubble time via GW emission. Compared to the case of BH companions, higher fractional mergers are formed for NS companions, because of the tighter orbits. For example, the GW merger fraction of $2.80\,\mathit M_{\odot}$ + NS is $44\%$, which is higher than the $9\%$ in the case of BH companion.

\section{CBD-binary interactions}
\label{CBD-binary interactions}
Based on the outcome of the 3D CE simulations, approximately $0.7\,\mathit M_{\odot}$ and $0.2\,\mathit M_{\odot}$ of material is gravitationally bound by the inner post-CE binary with a BH and NS companion, respectively. This material is expected to form a CBD surrounding the post-CE system (see Sect.\,\ref{Arepo_method}). As illustrated in Fig.\,\ref{fig:Disk_cartoon}, our study explores two distinct mechanisms through which the CBD interacts with the inner binary. On the one hand, mass and angular momentum may be accreted from the CBD onto the binary, leading to orbital expansion (see Sect.\,\ref{Accretion from disk_method}). On the other hand, a negative gravitational torque between the CBD and binary acts on the binary orbit via resonant interactions. This negative torque extracts orbital angular momentum from the binary to the CBD, thereby shrinking the binary orbit (see Sect.\,\ref{Resonant_method}). In this section, we will show how these different mechanisms impact the evolution of our post-CE binary systems without involving any effects from binary MT, with a particular emphasis on the fiducial RSG remnant of $2.80\,\mathit M_{\odot}$. The effects of CBD-binary interactions on the post-CE binaries with different RSG remnants of $2.85\,\mathit M_{\odot}$ and $2.75\,\mathit M_{\odot}$ are investigated in Appendix\,\ref{CBD-binary on other models}.

\subsection{Fiducial model with a BH companion}
 
We develop a grid of disk models with varying lifetimes ranging from $10^{2} \,\rm yr$ to $10^{5}\,\rm yr$, and masses in the range of $10^{-4} - 0.7\,\mathit M_{\odot}$ (as detailed in Sect.\,\ref{CBD_methods}). In this section, we will explore the influences of mass and angular momentum accretion and resonant interactions separately in Sect.\,\ref{Accretion from Circumbinary disk for BH} and Sect.\,\ref{Resonant interaction for BH}. The overall outcome induced by these two mechanisms is investigated in Sect.\,\ref{overall effects of CBD for BH}.

\subsubsection{Mass and angular momentum accretion from the CBD}
\label{Accretion from Circumbinary disk for BH}

\begin{figure}
\begin{center}
\includegraphics[width=0.5\textwidth]{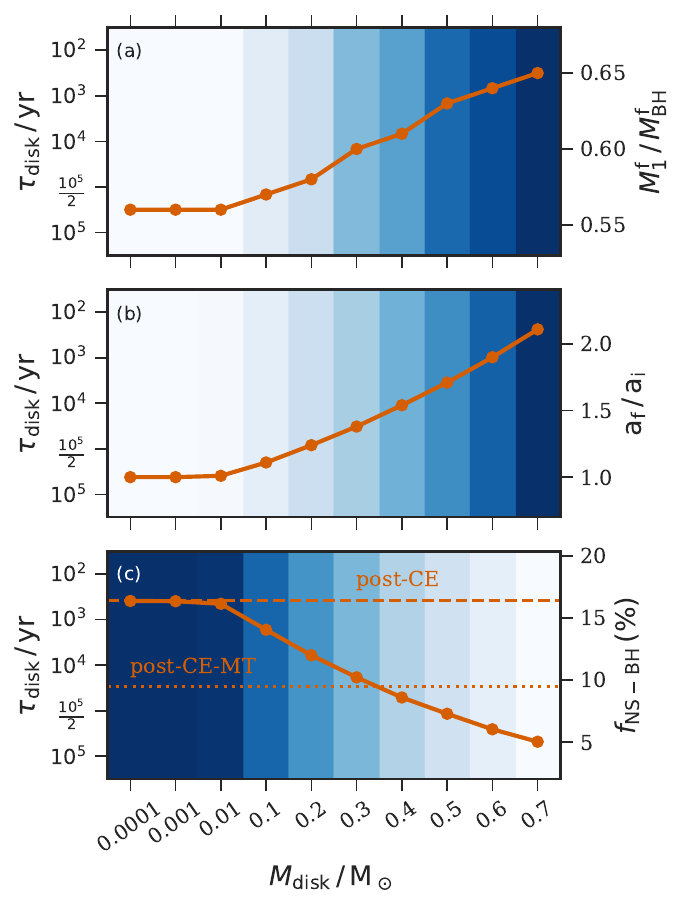}
\caption{Final mass ratio, final separation with respect to its initial value, and GW merger fraction after the SN explosion for $2.80\,\mathit M_{\odot}$ + BH binary systems after mass accretion from CBD models with different masses (bottom scale) and lifetimes (left scale). The darker blue colour indicates higher value, which is quantified by orange circle symbols (right scale). For comparison, the dashed orange line in panel c indicates the NS-BH merger fraction of ${\sim}\, 16\%$ for the post-CE binary without any interactions, and the merger fraction of ${\sim}\, 9\%$ for the post-CE-MT binary is denoted by the dotted line. }
\label{fig:only_acc}
   \end{center}
\end{figure}

As shown in Fig.\,\ref{fig:only_acc}, the accretion of CBD material reveals notable impacts on the post-CE binary of the $2.8\,\mathit M_{\odot}$ RSG remnant and the $5.0\,\mathit M_{\odot}$ BH companion. The accretion process results in an increase of the mass ratio $ M_{\rm 1}^{\rm f}\,/\,M_{\rm BH}^{\rm f}$, and of the orbital separation. Consequently, mass and angular momentum accretion from the CBD decreases the subsequent double compact merger fraction. 

Mass accretion from the CBD increases the mass of the RSG remnant, while a negligible amount of material is accreted by the compact object because of the Eddington-limited accretion. Consequently, the mass ratio ($ M_{\rm 1}^{\rm f}\,/\,M_{\rm BH}^{\rm f}$) approaches 1.0 as the CBD mass grows (see Fig.\,\ref{fig:only_acc}{\color{blue}{a}}), which is in line with the outcome of 2D hydrodynamical simulations from \citet{2023MNRASSiweka}. Given a fixed disk mass, CBDs with varying lifetimes exhibit consistent behaviour in terms of variations of the binary configuration (including mass ratio, separation, and double compact merger fraction). This consistency across different lifetimes is because of a significantly higher accretion rate from the CBD (${\ge}\,10^{-6}\,\mathrm{ M_{\odot}\, yr^{-1}}$ for CBD mass greater than $0.01\,\mathit M_{\odot}$) in comparison to the Eddington-limited accretion rate of $1.9\times10^{-7}\,\mathrm{M_{\odot}\, yr^{-1}}$. Hence, the same final mass ratio, separation and merger fraction are observed in binaries with CBDs of fixed mass, but varying lifetimes.

As the mass of the CBD increases, the post-CE-accretion orbital separation increases because of more angular momentum gained by the binary system (see Fig.\,\ref{fig:only_acc}{\color{blue}{b}}). The final orbit can reach approximately twice its initial value for the $0.7\,\mathit M_{\odot}$ CBD. Little material accretion from less massive disks ($10^{-4} - 10^{-2}\,\mathit M_{\odot}$) hardly influences both the post-CE-accretion binary configurations (see Figs.\,\ref{fig:only_acc}{\color{blue}{a}} and \ref{fig:only_acc}{\color{blue}{b}}) and the GW merger fractions (see Fig.\,\ref{fig:only_acc}{\color{blue}{c}}).

The final double compact merger fractions of post-CE binaries that experienced accretion from CBDs are shown in Fig.\,\ref{fig:only_acc}{\color{blue}{c}}. These fractions are obtained using the formalism described in Sect.\,\ref{GW merger rate for MT}. The double compact merger fraction decreases with higher disk masses because of the correspondingly wider orbit as shown in Fig.\,\ref{fig:only_acc}{\color{blue}{b}}. For comparison, the double compact merger fraction of the post-CE-MT binary is marked by the dotted line, which is half of that observed in the post-CE binary without any binary MT or CBD-binary interactions (dashed orange line in Fig.\,\ref{fig:only_acc}{\color{blue}{c}}). This is because the MT phase widens the orbit in the case of a BH companion as shown in Sect.\,\ref{Two fiducial models}. With the most massive CBD of $0.7\, \mathit M_{\odot}$, the compact merger fraction is reduced by a factor of 3 compared to that of the post-CE binary. So mass accretion from a CBD might modify the orbit more than the binary MT phase, and hence affect the final merger fractions more significantly.

\subsubsection{Resonant interactions}
\label{Resonant interaction for BH}

\begin{figure}
\begin{center}
\includegraphics[width=0.5\textwidth]{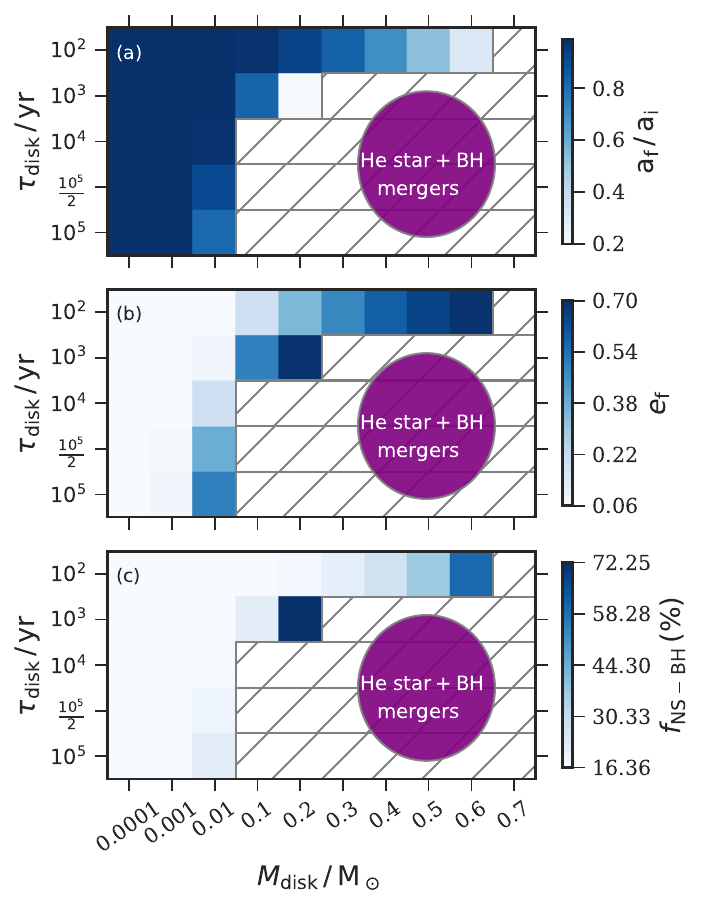}
\caption{Parameter space of the CBD lifetime as a function of the CBD mass. Colours indicate final separation with respect to the initial value, final eccentricity, and GW merger fraction for $2.80\,\mathit M_{\odot}$ + BH binary systems after the resonant interactions between the binary and different CBD models. Hatched grey regions denote the mergers induced by resonant interaction before the formation of the second compact object. As shown in Fig.\,\ref{fig:only_acc}, the double compact merger fraction of $2.80\,\mathit M_{\odot}$ + BH without any interactions is ${\sim}\, 16\%$, and around $9\%$ of binaries merge within a Hubble time after binary MT.}
\label{fig:only_RI}
   \end{center}
\end{figure}

Resonant interactions between the binary and CBD can shrink the binary orbit and pump the eccentricity, hence ultimately increasing the double compact merger rate (Fig.\,\ref{fig:only_RI}). For a more massive disk with a longer lifetime, the influence of resonant interaction between the binary and CBD is stronger, resulting in a substantial shortening of the binary orbit (Fig.\,\ref{fig:only_RI}{\color{blue}{a}}), even causing He star + BH mergers. In our models, we assume that the He star can merge with its compact companion as soon as the binary separation is shorter than the tidal radius \citep{1958ZAvonHoerner} of the He star given by
\begin{equation} 
\label{eqs.23}
r_{\rm tide} = \left( \frac{2M_{\rm BH\ or\ NS}}{M_{\rm He}} \right)^{\frac{1}{3}}R_{\rm He}.
\end{equation}
Here $M_{\rm He}$ and $M_{\rm BH\ or\ NS}$ are the masses of the He star with a radius of $R_{\rm He}$ and its compact companion, respectively. 

The rate of change of orbital separation due to resonant interactions is proportional to the mass ratio of the CBD and the binary (as described by Eq.\,\eqref{eqs.14}). That is, for a fixed binary configuration, a more massive disk induces stronger resonant interactions. For a fixed CBD mass, a longer lifetime means a smaller mass loss rate from the CBD in our models (Sect.\,\ref{CBD_methods}). As a result, a larger fraction of material in the CBD can interact with the inner binary over time, leading to more orbital shrinkage than in the case of a shorter lifetime. Certain He star + BH binary systems are found to merge prior to the second SN explosion because of the significant contraction of the orbital separation as indicated by the grey hatched regions in Fig.\,\ref{fig:only_RI}. He star + BH mergers occur for CBD masses exceeding $0.1\,\mathit M_{\odot}$. In the case of a shorter lifetime, the He star + BH merger can only take place for the massive CBD (i.e. CBD with $\tau_{\rm D} =10^2\,\rm yr$ and $M_{\rm D} = 0.7\,\mathit M_{\odot}$). However, for CBDs with a lifetime langer than $10^4\,\rm yr$, all of the models can induce He star + BH mergers.

The eccentricity of the post-CE binary, initially 0.06, increases when the CBD has a mass larger than $0.01\,\mathit M_{\odot}$. An equilibrium value of 0.7 (Sect.\,\ref{Resonant_method}) for the eccentricity is reached in some CBD models (i.e. CBD with $M_{\rm D}=0.2\,\mathit M_{\odot}$ and $\tau_{\rm D}=10^3\,\rm yr$, CBD with $M_{\rm D}=0.6\,\mathit M_{\odot}$ and $\tau_{\rm D}=10^2\,\rm yr$, and all the progenitors of the He star + BH mergers). 

Double compact mergers are possible via GW emission for post-CE binaries that do not merge before the second SN explosion. Influenced by the resonant interaction between the CBD and binary, the rates of GW-induced mergers within a Hubble time can increase by up to 5 times that of the post-CE binary without any interactions (including a MT phase and CBD-binary interaction). Conversely, the merger rate is decreased to half of that of the post-CE binary, after the binary MT (Fig.\,\ref{fig:only_acc}{\color{blue}{c}}). We find that the resonant interactions between the CBD and binary in the case of BH companion play a much more important role in facilitating the later double compact merger rate than the binary MT phase.

\subsubsection{Overall CBD-binary effects}
\label{overall effects of CBD for BH}

\begin{figure}
\begin{center}
\includegraphics[width=0.5\textwidth]{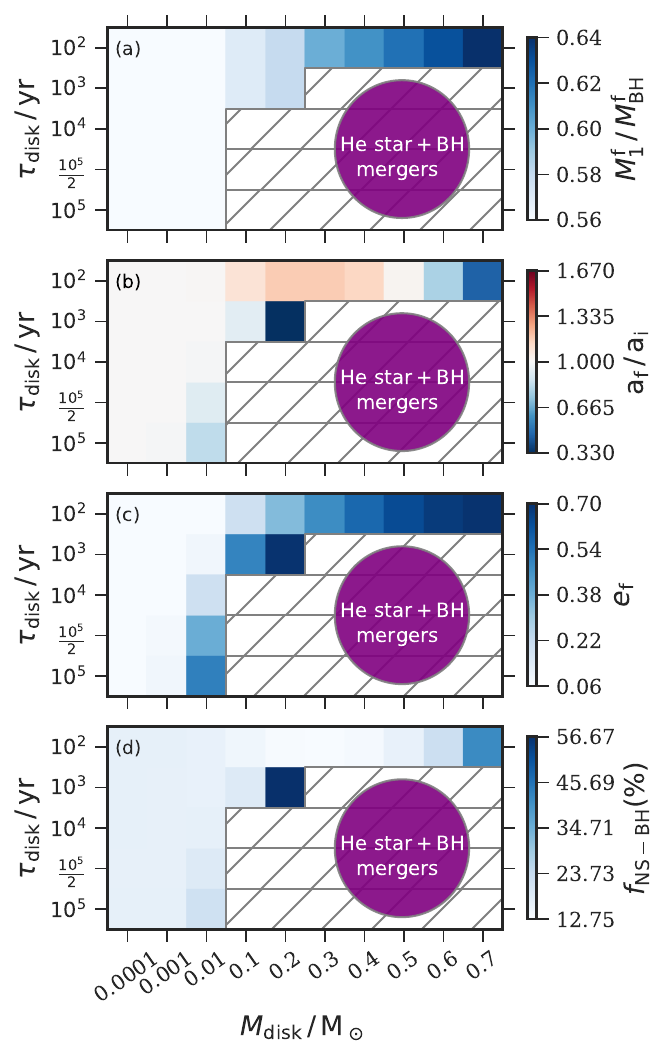}
\caption{Parameter space of the CBD lifetime as a function of the CBD mass. Colours indicate final mass ratio, ratio of final to initial orbital separation, final eccentricity, and GW merger fraction for $2.80\,\mathit M_{\odot}$ + BH binary systems after disk accretion and resonant interactions. Hatched grey regions denote mergers induced by the overall CBD-binary effects before the formation of the second compact object. As shown in Fig.\,\ref{fig:only_acc}, the double compact merger fraction of $2.80\,\mathit M_{\odot}$ + BH without any interactions is ${\sim}\, 16\%$, and around $9\%$ of binaries merge within a Hubble time after binary MT.}
\label{fig:disk_model_all_BH}
   \end{center}
\end{figure}

The overall effect of the CBD on the evolution of the post-CE binary depends on the competing effects between mass accretion from the disk and the resonant interactions. The changes of the binary mass ratio shown in Fig.\,\ref{fig:disk_model_all_BH}{\color{blue}{a}} are the same as those in the models only considering disk accretion onto a circular binary (c.f. Fig.\,\ref{fig:only_acc}). This is because in our binary model with the mass ratio of ${\sim}\, 0.5$, the fraction of mass accretion on different components does not depend much on the eccentricity \citep{2023MNRASSiweka}.

The orbital separation expands in certain binary systems with lifetimes ${\le}\,10^{2}\,\rm yr$, indicated by the red colour in Fig.\,\ref{fig:disk_model_all_BH}{\color{blue}{b}}. For instance, the orbit widens by one-fifth of its initial value in the case of a $0.3\,\mathit M_{\odot}$ CBD model with a lifetime of $10^{2}\,\rm yr$. In contrast, the $0.3\,\mathit M_{\odot}$ CBD model with a longer lifetime (e.g. ${\ge}\,10^{3}\,\rm yr$) leads to a contraction of the binary orbit. This is because the resonant interactions become stronger for the CBD with a longer lifetime, thus counteracting the orbital expansion from accretion. Moreover, the resonant interaction becomes stronger with larger CBD masses. Consequently, the binary orbit contracts for the more massive disk.
For most of our models, the orbits shrink (Fig.\,\ref{fig:disk_model_all_BH}{\color{blue}{b}}), in some cases even rapidly enough to cause He star + BH mergers before the second SN explosion.

The eccentricities are identical to those in models only taking into account the resonant interactions (cf. Fig.\,\ref{fig:only_RI}). Mergers before the formation of the second compact object are located in the similar mass-lifetime parameter space as in Fig.\,\ref{fig:only_RI}, with the exception of the model with a mass of $ 0.7\,\mathit M_{\odot}$ and a lifetime of $10^{2}\,\rm yr$, where angular momentum accretion from the CBD helps this model to avoid merging. 

The resulting double compact merger fractions after the overall CBD-binary interactions decrease at most to $12\%$, or rise up to ${\sim}\, 56\%$, compared to the post-CE binary without any interactions. As discussed in Sect.\,\ref{GW merger rate for MT}, in the case of a BH companion, the binary MT decreases the merger fraction of the post-CE binary from ${\sim}\, 16\%$ to ${\sim}\, 9\%$. We find that in the case of a post-CE binary with a BH companion, the CBD-binary interactions may be crucial in promoting the merger fraction of later double compact objects, while the additional binary MT has the opposite effect (Sect.\ref{GW merger rate for MT}).

\subsection{Fiducial model with a NS companion}

For the post-CE binary with a NS companion, we consider CBDs with masses of $10^{-4} - 0.2\,\mathit M_{\odot}$, and lifetimes of $10^{2} - 10^{5}\,\rm yr$. The maximum CBD mass of $0.2\,\mathit M_{\odot}$ is less massive than that in the case of BH companion because of the deeper spiral-in, according to the outcome of the 3D CE simulations (Sect.\,\ref{Arepo_method}). In this section, we demonstrate how the CBDs affect the post-CE evolution with a NS companion and the final double compact merger fraction and compare to the case of a BH companion.

\subsubsection{Mass and angular momentum accretion from the CBD}
\begin{figure}
\begin{center}
\includegraphics[width=0.5\textwidth]{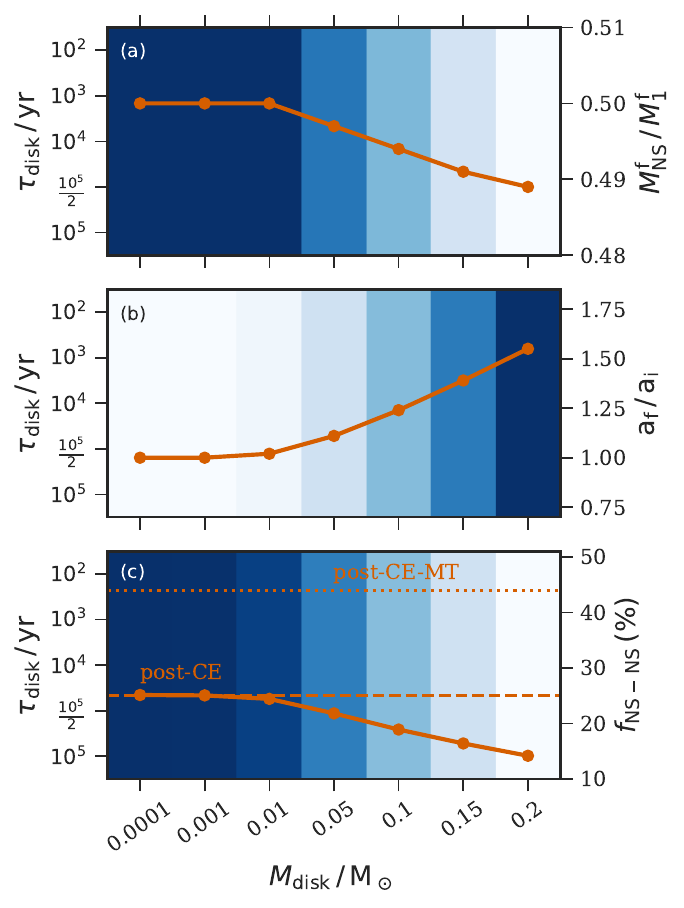}
\caption{Same as Fig.\,\ref{fig:only_acc} but for $2.80\,\mathit M_{\odot}$ + NS binary systems. For comparison, the dashed orange line in panel c indicates the NS-NS merger fraction of ${\sim}\, 25\%$ for the post-CE binary without any interactions, and around $44\%$ of binaries merge within a Hubble time after binary MT (post-CE-MT), which is denoted by the dotted orange line.}
\label{fig:only_acc_NS} 
   \end{center}
\end{figure}

Fig.\,\ref{fig:only_acc_NS} shows the impact of material accretion from the CBD on the evolution of the post-CE binary of a $2.8\,\mathit M_{\odot}$ RSG remnant orbiting a $1.4\,\mathit M_{\odot}$ NS companion. The CBD mass around the binary of $2.80\,\mathit M_{\odot}$ + NS reaches $0.2\,\mathit M_{\odot}$ at most, which is smaller than that in $2.80\,\mathit M_{\odot}$ + BH binaries (Fig.\,\ref{fig:only_acc}). Therefore, also the variation of the mass ratio is smaller in case of the NS companion (Fig.\,\ref{fig:only_acc_NS}{\color{blue}{a}}) compared to that shown in Fig.\,\ref{fig:only_acc}{\color{blue}{a}}. The mass ratio ($M_{\rm NS}^{\rm f}\,/\,M_{\rm 1}^{\rm f}$) decreases with increasing disk mass, because the accretion onto the NS companion is Eddington limited, while the RSG remnant accretes material following the preferential accretion model obtained from 2D-CBD simulations.

Similar to the model with a BH companion, mass accretion increases the binary separation because of the angular momentum accretion from the disk (Fig.\,\ref{fig:only_acc_NS}{\color{blue}{b}}), and the separation becomes wider as the CBD disk mass increases. In addition, for the NS companion, the binary separation exhibits slightly more expansion than in the case of the BH companion for the same CBD model. For example, for a CBD mass of $0.2\,\mathit M_{\odot}$, the separation increases by a factor of 1.55 for the NS companion, while it does by a factor of 1.24 for the BH companion (Fig.\,\ref{fig:only_acc}{\color{blue}{c}}). Notably, in the case of NS companion, the binary interaction shown in Sect.\,\ref{Additional mass transfer} decreases the binary separation, while mass accretion from the disk works in the opposite direction.

Fig.\,\ref{fig:only_acc_NS}{\color{blue}{c}} shows the final double compact merger fraction within a Hubble time for the post-CE, post-CE-MT and post-CE-accretion binary. The merger fraction of the post-CE binary ($2.80\,\mathit M_{\odot}$ + NS), without any binary MT or CBD-binary interaction, is ${\sim}\, 25\%$ as indicated by a dashed line. After the binary MT as shown in Sect.\,\ref{Additional mass transfer}, over $40\%$ of the post-CE-MT binary can merge. It is because, in the case of NS companion, the binary MT decreases the binary separation. However, the wider separation due to the angular momentum accretion from the CBD decreases the double compact merger fraction by a factor of 1.6 for the $0.2\,\mathit M_{\odot}$ CBD, compared to the post-CE binary without any interactions.

\subsubsection{Resonant interactions}
\begin{figure}
\begin{center}
\includegraphics[width=0.5\textwidth]{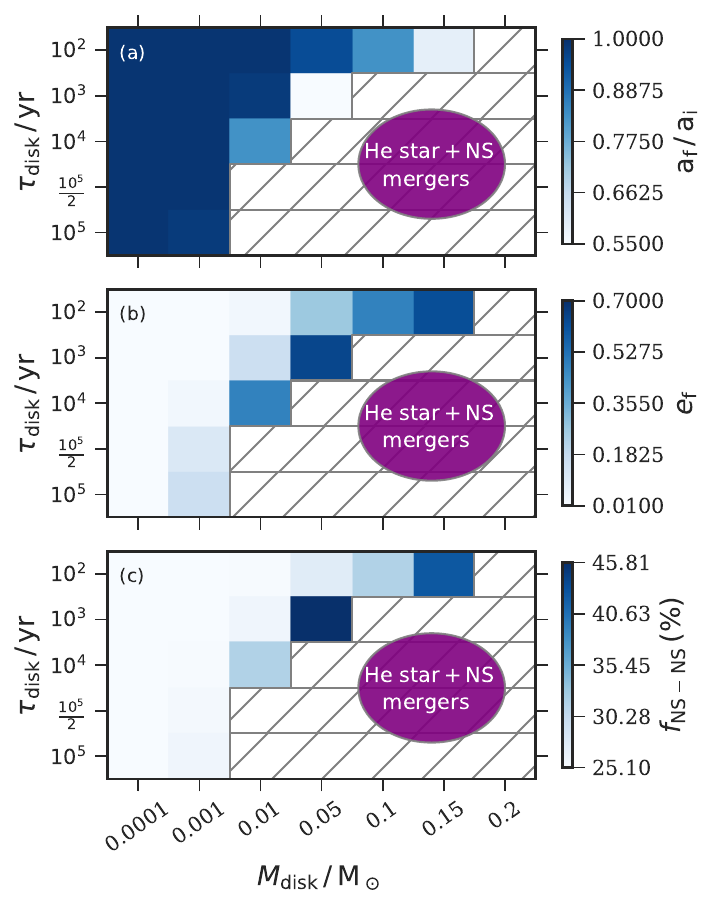}
\caption{Same as Fig.\,\ref{fig:only_RI} but for $2.80\,\mathit M_{\odot}$ + NS binary systems. As shown in Fig.\,\ref{fig:only_acc_NS}, the double compact merger fraction of $2.80\,\mathit M_{\odot}$ + BH without any interactions is ${\sim}\, 25\%$, and around $44\%$ of binaries merge within a Hubble time after binary MT.}
\label{fig:only_RI_NS}
   \end{center}
\end{figure}

As shown in Fig.\,\ref{fig:only_RI_NS}, the resonant interaction between the CBD and post-CE binary decreases the binary separation (Fig.\,\ref{fig:only_RI_NS}{\color{blue}{a}}), increases the eccentricity (Fig.\,\ref{fig:only_RI_NS}{\color{blue}{b}}) and hence increases the final double compact merger fraction (Fig.\,\ref{fig:only_RI_NS}{\color{blue}{c}}). Effects of the increased eccentricity on the post-SN binary configurations are investigated in Appendix\,\ref{effects of increased eccentricity on the post-SN binary configurations}. A stronger resonant interaction is produced by a more massive CBD with a longer lifetime. Because of the more efficient separation contraction, most He stars are found to merge with their NS companions before the second SNe explosion, when the post-CE binary has a CBD with $M_{\rm D} {\ge}\,0.01\,\mathit M_{\odot}$ and $\tau_{\rm D} {\ge}\, 10^{3}\,\rm yr$. Compared to the case of a BH companion, the post-CE binary orbit of the $2.80\,\mathit M_{\odot}$ + NS binary is tighter (see Sect.\,\ref{post-CE_method}), leading to stronger resonant interaction for a given CBD model (Fig.\,\ref{fig:only_RI_NS}). For instance, given a CBD with $M_{\rm D} = 0.2\,\mathit M_{\odot}$ and lifetime of $\tau_{\rm D} = 10^{2}\,\rm yr$, the He star in the post-CE binary with a NS companion has the potential to merge with its compact companion. However, in the case of a BH companion, the post-CE binary survives from the resonant interactions with the final eccentricity of 0.35, and the final separation with respect to its initial value is 0.93.

For post-CE binaries that do not produce He star + NS mergers, the final double compact merger fractions after the second SN explosion are shown in Fig.\,\ref{fig:only_RI_NS}{\color{blue}{c}}. With a CBD of $M_{\rm D} = 0.05\,\mathit M_{\odot}$ and $\tau_{\rm D} = 10^{3}\,\rm yr$, over $45\%$ of post-CE binaries can produce GW-induced double compact mergers within a Hubble time, which is almost 2 times the merger fraction for post-CE binary without any interactions (${\sim}\, 25\%$), and also slightly higher than the post-CE-MT binary (${\sim}\, 44\%$). We find that the resonant interactions might be as efficient as the binary MT in producing GW-induced mergers for the post-CE binary with a NS companion.

\subsubsection{Overall CBD-binary effects}
\begin{figure}
\begin{center}
\includegraphics[width=0.5\textwidth]{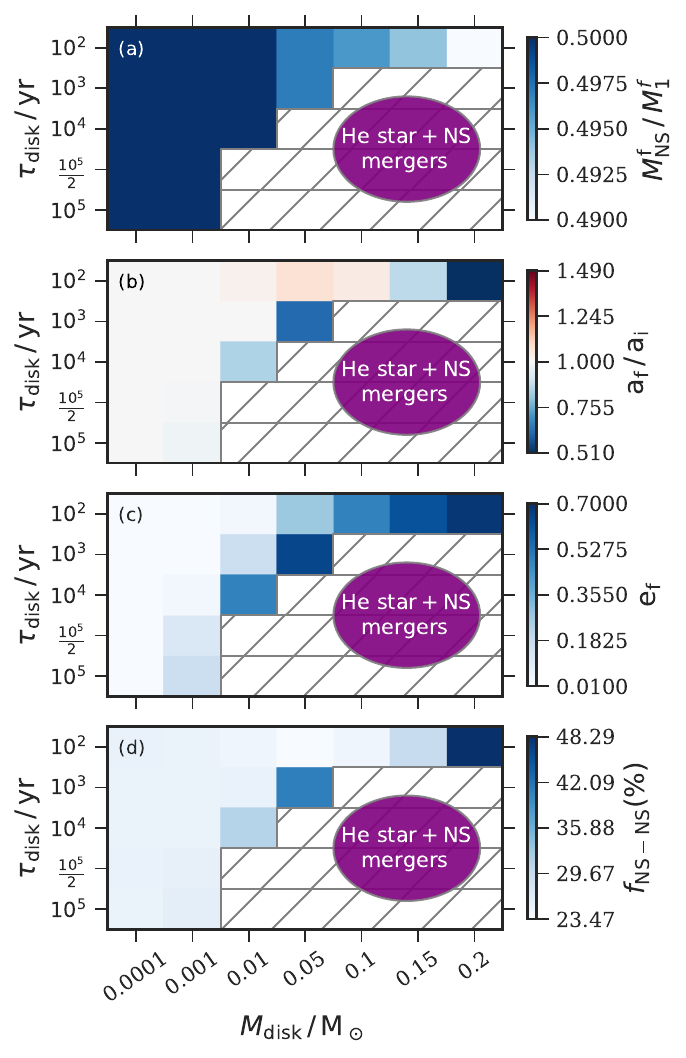}
\caption{Same as Fig.\,\ref{fig:disk_model_all_BH} but for $2.80\,\mathit M_{\odot}$ + NS binary systems. As shown in Fig.\,\ref{fig:only_acc_NS}, the double compact merger fraction of $2.80\,\mathit M_{\odot}$ + BH without any interactions is ${\sim}\, 25\%$, and around $44\%$ of binaries merge within a Hubble time after binary MT.}
\label{fig:disk_model_all_NS}
   \end{center}
\end{figure}

The overall effect of CBD-binary interactions in the case of a NS companion, including the effects of the mass accretion and resonant interactions, is shown in Fig.\,\ref{fig:disk_model_all_NS}. Similar to the case of a BH companion, the changes of mass ratio in Fig.\,\ref{fig:disk_model_all_NS}{\color{blue}{a}} and eccentricity in Fig.\,\ref{fig:disk_model_all_NS}{\color{blue}{c}} are the same as that in Fig.\,\ref{fig:only_acc_NS}{\color{blue}{a}} and Fig.\,\ref{fig:only_RI_NS}{\color{blue}{b}}, respectively. As shown in Fig.\,\ref{fig:disk_model_all_NS}{\color{blue}{b}}, certain post-CE binaries have wider separations because the effect of mass accretion is stronger than the resonant interaction, i.e. CBD with $\tau_{\rm D} = 10^{2}\,\rm yr$ and $ 0.01\,{\le}\,M_{\rm D} {\le}\, 0.1 \,\mathit M_{\odot}$. However, the overall effect of CBD-binary interactions is to shrink the orbital separation of the post-CE binary in most CBD models, and these interactions can even lead to He star + NS mergers for certain CBDs with $\tau_{\rm D} {\ge}\, 10^{3}\,\rm yr$ and $M_{\rm D} {\ge}\,0.01\,\mathit M_{\odot}$.

Compared to the post-CE binary with a BH companion (Fig.\,\ref{fig:disk_model_all_BH}), for a given CBD, the overall effect of CBD-binary interactions in the case of NS companion works more efficiently in shrinking binary orbits, increasing eccentricity (Fig.\,\ref{fig:disk_model_all_NS}). This difference can, e.g., be observed for a CBD with a mass of $0.2\,\mathit M_{\odot}$ and lifetime of $10^{2}\,\rm yr$. In the post-CE binary with a NS companion, the separation decreases to 0.51 of its initial value, accompanied by an increase in eccentricity to nearly 0.7. For the BH companion, the impact of resonant interactions is comparatively weaker than mass accretion from the CBD. Therefore, the binary orbit expands with the eccentricity reaching roughly 0.34.

Fig.\,\ref{fig:disk_model_all_NS}{\color{blue}{d}} shows the final double compact merger factions for post-CE binaries which do not experience He star + NS mergers before the second SN explosion. Under the impact of CBD-binary interaction in the case of CBD with $\tau_{\rm D} = 10^{2}\,\rm yr$ and $M_{\rm D} = 0.2\,\mathit M_{\odot}$, a slightly higher fraction of ${\sim}\, 48\%$ of double compact objects can merge within a Hubble time compared to the post-CE-MT binaries. It is notable that the CBD-binary interactions might play an as important role as the binary MT on the evolution and final fate of the post-CE binaries with a NS companion.

\section{Discussion}
\label{Discussion}

\subsection{Additional MT considerations}
Binary interaction occurring at different evolutionary stages plays a crucial role in determining the final CO core mass prior to core collapse and consequently influences the type of compact remnant left behind \citep{2004ApJPodsiadlowski,2015MNRASTauris}. In this work, the RSG stars initiate the CE phase after core He depletion, leaving behind a CO core with a He envelope. Subsequently, Case BB MT develops after the CE phase. The majority of our models are expected to explode as Ib/c CCSNe, rather than ultra-stripped SNe, because the remnant He-rich envelope of our pre-SN models has a mass exceeding $0.2\,\mathit M_{\odot}$. However, these models have the potential to give rise to ECSNe, or result in a massive ONeMg WD, if the preceding RSG binaries initiate the CE before the core helium depletion since less massive He core size can be produced in this circumstance.

Several studies of helium stars + NS (or BH) binary evolution in tight orbits have been done to investigate the formation of double compact merger events \citep{2013ApJTauris,2015MNRASTauris}. Their models (called T models) begin as pure core-helium-burning stars and do not include the CO core structure initially, while the evolved He stars in our model (called evolved He models) are post-helium-burning stars with a CO core and helium envelope. Given the similar binary configurations, the T model in \citet{2013ApJTauris,2015MNRASTauris} loses more envelope mass during the evolution than the evolved He model, and possibly explodes in an ultra-stripped SN. To understand the differences between the evolution of the T model and the evolved He model during Case BB MT, we conduct a simulation involving a helium star of $2.80\,\mathit M_{\odot}$ (similar to the T model) orbiting a $1.40\,\mathit M_{\odot}$ NS with the same orbital period of $0.53\,\rm d$ as that in our fiducial evolved He model with a NS companion. This helium star initiates MT to its companion during core C burning. We find that the helium star model forms a smaller final CO core mass ($M_{\rm core,f}^{\rm CO} {\sim}\, 1.43\,\mathit M_{\odot}$), and a smaller He-envelope ($M_{\rm He,f}^{\rm env} {\sim}\, 0.12\,\mathit M_{\odot}$) prior to collapse, in comparison to the evolved He model in our fiducial NS model ($M_{\rm core,f}^{\rm CO} {\sim}\, 1.56 \,\mathit M_{\odot}$, $M_{\rm He,f}^{\rm env} {\sim}\, 0.23\,\mathit M_{\odot}$ ). This difference arises because the wind mass loss during the core He burning strips away around $0.3\,\mathit M_{\odot}$ of the envelope in the He star, resulting in a less massive final CO core and envelope, and an ultra-stripped SNe. However, our evolved He model is post-helium-burning, and the wind mass loss during the core He burning does not affect its evolution. Therefore, we find that for the case BB MT occurring after the CE phase of a post-helium-burning RSG, the pure helium model computed with the assumption of wind mass loss may underestimate the mass of the remaining envelope, compared to our evolved He model in the fixed binary configuration. We note that the wind mass loss rates of stripped stars are still very uncertain. Recent theoretical work \citep{2020MNRASSander} and observations \citep{2023SciDrout} suggest that it is likely significantly lower than assumed in our models, which would potentially lead to an even larger remaining He-rich envelope in both our evolved He models and the T models.

\subsection{CBD considerations}

Variations in CBD parameters, such as viscosity ($\alpha$), density parameter ($\delta$) and aspect ratio, introduce uncertainties that can yield different effects on the post-CE binary evolution, and also on the final double compact merger fraction within a Hubble time. To investigate the effects of different parameters, we utilise the binary model of $2.80\,\mathit M_{\odot}$ + NS surrounding by a $0.05\,\mathit M_{\odot}$ CBD with a lifetime of ${\sim}\, 10^3\,\rm yr$ as a benchmark model. All of the parameters describing this CBD are originally defined in Sect.\,\ref{CBD_methods}. 

The viscosity parameter $\alpha$ may span a range from $10^{-3}$ to $10^{-1}$, as in observations of protoplanetary disks \citep{2017ApJRafikov,2018ApJAnsdell}. For the benchmark model, we modify the viscosity parameter $\alpha$ from 0.1 to 0.001. This results in a dampening of the resonant interaction, leading to a reduction of the merger fraction to 21\% for the double compact binary. The merger fraction of the benchmark model with $\alpha = 0.1$ is around 40\%.

The density distribution of the CBD has a crucial impact on the CBD-binary interactions; however, it is an uncertain physics process affected by the disk evolution. In our models, the density distribution is simplified to be constrained by the time-dependent angular momentum and mass of the CBD, modulated by the parameter $\delta$. Reducing $\delta$ from 1.5 to 1.0 weakens the resonant interaction between the binary and CBD, inducing a smaller double compact merger fraction (${\sim}\, 22\% $) within a Hubble time.

The nature of jet-like outflows during the CE phase is still a puzzling question \citep{2015ApJSoker,2018arXivVan, 2019MNRASSoker,2022A&AOndratschek,2023LRCAFritz}. These outflows have the potential to expel material from the CBD, reducing the CBD mass, even after the dynamical spiral-in phase of the CE phase. This mass alteration in the CBD can significantly impact the subsequent evolution of the inner post-CE binary as shown in Fig.\,\ref{fig:disk_model_all_BH} and Fig.\,\ref{fig:disk_model_all_NS}. Moreover, the angular momentum in the CBD may be diminished because of the outflows, potentially facilitating a double compact merger event. For the benchmark model, ${\sim}\, 40\%$ of NS-NS binaries can merge within a Hubble time, if the angular momentum accretion rate ($\dot{J}_{\rm acc}$) follows Eq.\,\eqref{eqs.10}. However, given the effect of outflows, if we modify the angular momentum accretion rate to $10\%$ of $\dot{J}_{\rm acc}$, a substantial contraction in the binary orbit yields an enhanced double compact merger fraction of ${\sim}\, 46\%$. These findings demonstrate that the jet-like outflows observed in the 3D CE simulations (\citet{2022A&AMoreno} and Vetter et al. (in prep)) may play a crucial role in the evolution of post-CE binary interaction.

CBD lifetimes play an important role in the CBD-binary interactions (see Sect.\,\ref{CBD-binary interactions}), which can be comparable to the thermal timescale of the remnant He star and even reach its nuclear timescale \citep{2023ApJTuna}. Given the long lifetime, the impact of CBD-binary interactions on the post-CE binary evolution can be even more important than the binary MT. From Figs.\,\ref{fig:disk_model_all_BH} and \,\ref{fig:disk_model_all_NS}, the disk models lasting for a thermal or nuclear timescale ($10^3 - 10^5\,\rm yr$) can significantly affect the configuration of post-CE binaries (e.g. orbit contraction, eccentricity pumping) via CBD-binary interactions. With the disk mass larger than $0.1\,\mathit M_{\odot}$, the orbital contraction due to resonant interactions is efficient enough to lead to He star + BH mergers in the case of a BH companion. For the NS companion in Fig.\,\ref{fig:disk_model_all_NS}, CBDs with a mass exceeding $0.01\,\mathit M_{\odot}$ can induce He star + NS mergers. 

According to the CBD hydrodynamic simulations in \citet{2023MNRASSiwekb}, binaries with mass ratio $q\,{\ge}\,0.3$ tend to evolve towards coalescence, once they reach the equilibrium eccentricity. In our models, the mass ratios of the post-CE binaries exceed 0.3, and all of the He star + BH (or NS) mergers achieve an equilibrium eccentricity of 0.7, consistent with the findings in CBD hydrodynamic simulations.

In this work, we investigate how the additional MT and CBD-binary interactions affect the evolution of post-CE binaries, separately. Stable thermal timescale MT episodes are found in our post-CE binaries (Sect.\,\ref{Additional mass transfer}). However, the influence of CBD-binary interactions onto the stability of the MT phase cannot be neglected, because of the comparable timescales \citep{2023ApJTuna}. For more massive disk models, the orbital shrinkage is faster because of CBD-binary interactions, which can result in a higher MT rate. Consequently, unstable binary MT may occur \citep{2020ApJGe,2020ApJSGe,2023ApJGe,2023MNRASLu,2023ApJTuna}. Moreover, the binary MT can also be affected by the pumped eccentricity from CBD-binary interactions \citep{2016ApJDosopoulou_a,2016ApJDosopoulou_b}. Detailed models of the evolution of binaries with CBDs are essential in future work.

\subsection{GW mergers in a Hubble time}

In our fiducial models following the MT process, the GW-induced merger fractions (Table\,\ref{tab:pre-after-MT}) are quite similar to those for the post-CE binary with an orbital separation smaller by two-thirds in \citet{2022A&AMoreno}. This result is consistent with the fact that the separation of our fiducial model is nearly one-third of its original final separation obtained from 3D CE simulations, whether involving a BH or NS companion. 

The additional MT can strip the RSG remnant even further after the CE interaction, leading to a small iron core and less massive envelope ultimately. Consequently, a newly formed NS may receive a slower kick of several 10\,$\rm km\,s^{-1}$ after the SN explosion \citep{2015MNRASSuwa,2017ApJJanka}. According to the discussion in \citet{2022A&AMoreno}, a larger fraction of the binaries can remain bound after the SN explosion when a slower NS kick of 50\,$\rm km\,s^{-1}$ is imparted to the new-born NS. The merger fraction of the post-CE-MT binary with a shorter orbit (e.g. cases with NS companions) may increase, while it may decrease for the post-CE-MT binary with a wider orbit (e.g. cases with BH companions).

\subsection{Observations}
(Ultraluminous) X-ray emission may be observed both during the additional binary MT and the disk accretion phase. The mass accretion rates exceed the Eddington limit by $2{-}4$ orders of magnitude in our model. The non-accreted material could be expelled in the form of jet-like outflows \citep{2015MNRASSdowski,2016MNRASSdowski,2022ApJSridhar}.

If the CBD shrinks the binary separation sufficiently (see Fig.\,\ref{fig:disk_model_all_BH} and \ref{fig:disk_model_all_NS}), we find He star + NS (or BH) mergers. The He star + NS (or BH) merger process is not well understood so far. A 3D simulation of NS + COWD merger process in \citet{2023A&AMoran-Fraile} demonstrates that the COWD is tidally disrupted by the NS, resulting in an energetic transient. According to this simulation, we predict that the He star in our models may transfer mass to the compact companion when it overfills its Roche lobe radius, as the orbit shrinks because of the CBD-binary interactions. Then the compact companion will tidally disrupt the He star and accrete its mass at possibly high accretion rates. In addition, the magnetic fields can be amplified by this merger process, hence resulting in polar jets. During this high-rate accretion phase, an energetic explosion may be powered, such as a gamma-ray burst-like transient in the case of a BH companion \citep{1998ApJFryer,2023ApJTuna}. For a NS companion, an accretion-induced collapse may take place if the NS accreted enough material to reach the maximum NS mass, and a gamma-ray burst-like transient may be detected. Otherwise, the He star + NS merger event may resemble luminous fast blue optical transients if the NS does not collapse to a BH \citep{2020ApJSchroder,2022ApJMetzger,2023ApJTuna}. In addition, the merger object may become a Thorne-\.Zytkow object in the case of a merger with a NS companion \citep{1975ApJThorne,1977ApJThorne}. A detailed simulation of the He star + NS (or BH) merger needs to be conducted in future research.

The He stars can explode as Ib/c SNe in our models if the He star + NS (or BH) binaries survive the orbital contraction induced by the CBD-binary interactions. Furthermore, when the SN ejecta interacts with the pre-existing circumstellar material created during the binary interacting, an interacting SNe may be observed \citep{2022ApJWu,2024ApJMatsuoka,2024A&AErcolino}. More details about this topic will be presented in Wei et al. (in prep). After the second SN explosion, the GW emission is released during the double compact merger event.

\section{Conclusion}
\label{Concusion}

We investigated the evolution and final fate of massive post-CE binaries, based on the outcome of a 3D CE interaction of a $9.4\,\mathit M_{\odot}$ RSG and a compact companion (a $1.4\,\mathit M_{\odot}$ NS or $5.0\,\mathit M_{\odot}$ BH). In our work, we explore the effect of additional MT and CBD-binary interactions on the evolution of the massive post-CE binaries. Our results demonstrate that both the additional MT and the CBD-binary interactions can influence the orbital evolution of post-CE binaries, hence affecting the final fractions of double compact object mergers in a Hubble time. In addition, the CBD-binary interactions may be more crucial in facilitating the GW-induced mergers than the binary MT phase. 
Our main results can be summarised as follows.
\begin{itemize}
\item  As a response to rapid envelope removal, the core of the RSG expands, resulting in more mass being ejected from the system than the 3D CE models predicted.
\item MT episodes in the post-CE binaries can shape the final binary configurations and final stellar structures. Most of our models are expected to explode as Ib/c CCSNe, and only one post-CE binary with the shortest orbit ($2.75 \,\mathit M_{\odot}$ + NS) may explode as an ultra-stripped CCSN. 
\item Additional MT changes the orbital evolution of the post-CE binaries. In the case of a BH companion, additional MT increases the binary separation. In contrast, the binary orbit shrinks for the NS companion. 
\item In the post-MT binaries with BH companions, the orbit of double compact objects is too wide to merge in a Hubble time without NS kicks. NS kicks are essential in these cases to sufficiently perturb the orbit and thus facilitate mergers via GW emission.
\item Mass and angular momentum accretion from the CBD widens the binary, resulting in a smaller double compact merger fraction. Resonant interactions between the binary and CBD can shrink the binary orbit and pump the eccentricity, hence increasing the double compact merger fraction, and even leading to He star + NS and BH mergers in some cases. Many interesting astrophysical objects can potentially be observed during the He star + NS and BH merger process, for example, ultraluminous X-ray binary, gamma-ray bursts, and luminous fast blue optical transients.  
\item Under the overall CBD-binary effects, the post-CE binary orbit can expand or shrink, depending on the competition between CBD mass accretion and resonant interactions. Orbits shrink for most of our binary models with CBDs, increasing the subsequent double compact merger fraction.
\end{itemize}

We conclude that the CBD around the post-CE binary can alter the binary configuration and potentially increase the ultimate GW-induced mergers. The CBD-binary interactions might play an as important role as the binary MT on the evolution and final fate of the post-CE binaries, and may be crucial to better understand observed GW mergers.

\begin{acknowledgements} 
We thank the referee for their insightful comments which are very helpful to improve this work further. We thank Vincent Bronner, Jan Henneco and Heran Xiong for the meaningful discussion and suggestions. We are grateful for support from the Klaus Tschira Foundation. This work has received funding from the European Research Council under the European Union's Horizon 2020 research and innovation program (grant agreement No. 945806 and 101096243). This work is supported by the Deutsche Forschungsgemeinschaft (DFG, German Research Foundation) under Germany's Excellence Strategy EXC 2181/1-390900948 (the Heidelberg STRUCTURES Excellence Cluster). This work is also supported by the post-doctoral scholarship from the Chinese Scholarship Council, Deutscher Akademischer Austauschdienst and the National Natural Science Foundation of China (Nos. 12073070).
\end{acknowledgements}

\bibliographystyle{aa}
\bibliography{refs}

\clearpage

\begin{appendix}

\section{Overall CBD-binary effects on the post-CE binaries with different RSG remnant masses}
\label{CBD-binary on other models}

\subsection{BH companion cases}

\begin{figure}
\begin{center}
\includegraphics[width=0.5\textwidth]{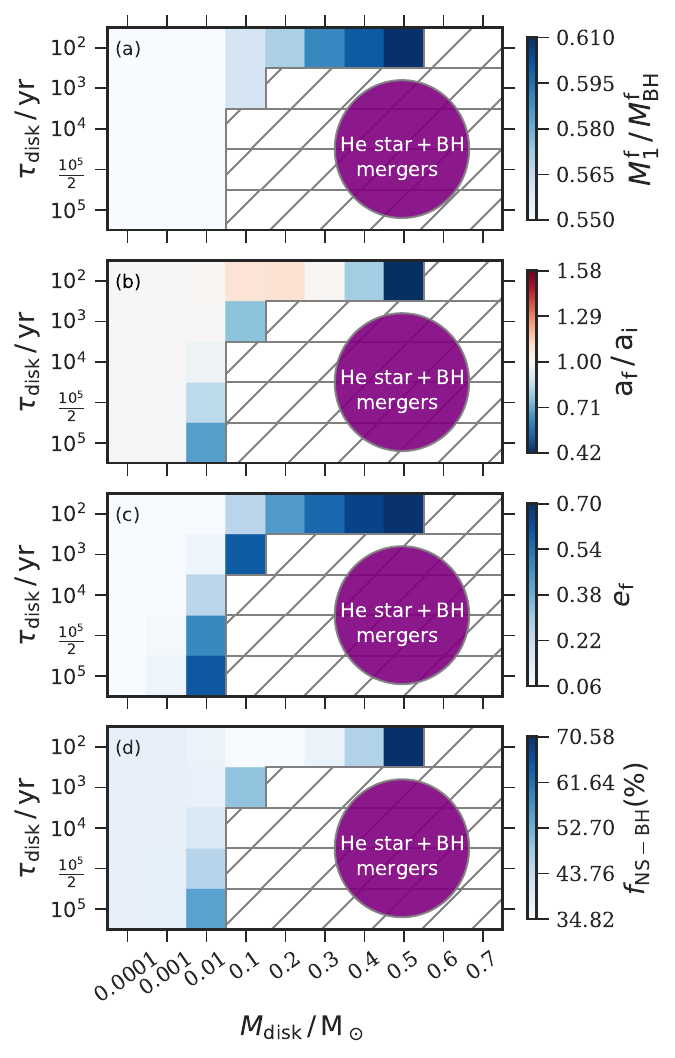}
\caption{Same as Fig.\,\ref{fig:disk_model_all_BH} but for the $2.75\,\mathit M_{\odot}$ + BH binary systems. For comparison, the double compact merger fraction of $2.75\,\mathit M_{\odot}$ + BH without any interactions is ${\sim}\, 37\%$, and around $21\%$ of binaries merge within a Hubble time after binary MT.}
\label{fig:disk_model_all_BH_2_75}
   \end{center}
\end{figure}

\begin{figure}
\begin{center}
\includegraphics[width=0.5\textwidth]{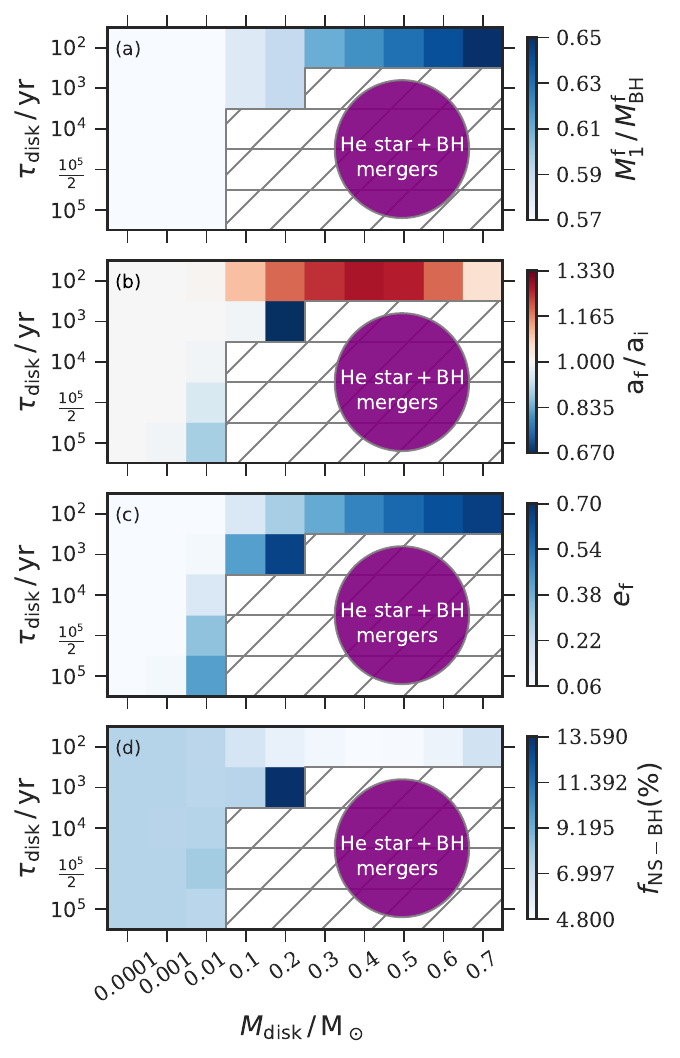}
\caption{Same as Fig.\,\ref{fig:disk_model_all_BH} but for the $2.85\,\mathit M_{\odot}$ + BH binary systems. For comparison, the double compact merger fraction of $2.85\,\mathit M_{\odot}$ + BH without any interactions is ${\sim}\, 7\%$, and around $5\%$ of binaries merge within a Hubble time after binary MT.}
\label{fig:disk_model_all_BH_2_85}
   \end{center}
\end{figure}

In the case of a BH companion, the dependence of CBD-binary interactions on the evolution of post-CE binaries with different RSG remnant masses ($2.75\,\mathit M_{\odot}$, $2.80\,\mathit M_{\odot}$ and $2.85\,\mathit M_{\odot}$) is shown in Fig.\,\ref{fig:disk_model_all_BH_2_75}, Fig.\,\ref{fig:disk_model_all_BH} and Fig.\,\ref{fig:disk_model_all_BH_2_85}, respectively. 
There are several commons on the outcomes of these post-CE binaries with a BH companion. The CBD-binary interactions, including the mass accretion and resonant interactions, can increase the binary mass ratio (Fig.\,\ref{fig:disk_model_all_BH_2_75}{\color{blue}{a}}, Fig.\,\ref{fig:disk_model_all_BH}{\color{blue}{a}} and Fig.\,\ref{fig:disk_model_all_BH_2_85}{\color{blue}{a}}) and eccentricity (Fig.\,\ref{fig:disk_model_all_BH_2_75}{\color{blue}{c}}, Fig.\,\ref{fig:disk_model_all_BH}{\color{blue}{c}} and Fig.\,\ref{fig:disk_model_all_BH_2_85}{\color{blue}{c}}). The orbital separation can be wider for several CBDs with a shorter lifetime (e.g. $\tau_{\rm D} = 10^{2}\,\rm yr$), and it can be tighter in most CBD models (Fig.\,\ref{fig:disk_model_all_BH_2_75}{\color{blue}{b}}, Fig.\,\ref{fig:disk_model_all_BH}{\color{blue}{b}} and Fig.\,\ref{fig:disk_model_all_BH_2_85}{\color{blue}{b}}). The different impacts on the binary separation depend on the competition between the two aspects in the CBD-binary interactions. 

Comparing Fig.\,\ref{fig:disk_model_all_BH_2_75}{\color{blue}{b}}, Fig.\,\ref{fig:disk_model_all_BH}{\color{blue}{b}} and Fig.\,\ref{fig:disk_model_all_BH_2_85}{\color{blue}{b}}, we find that in the case of a smaller RSG remnant, the binary contraction due to the resonant interaction is stronger than the case of a larger RSG remnant for a given CBD model. For instance, for a given CBD model with $\tau_{\rm D} = 10^{2}\,\rm yr$ and $M_{\rm D} = 0.5\,\mathit M_{\odot}$, the separation in the case of $2.75\,\mathit M_{\odot}$ + BH decreases by a factor of 0.42 (Fig.\,\ref{fig:disk_model_all_BH_2_75}{\color{blue}{b}}), while the orbits are wider in both cases of $2.80\,\mathit M_{\odot}$ + BH (Fig.\,\ref{fig:disk_model_all_BH}{\color{blue}{b}}) and $2.85\,\mathit M_{\odot}$ + BH (Fig.\,\ref{fig:disk_model_all_BH_2_85}{\color{blue}{b}}). These differences are driven by the differences in the initial separations of post-CE binaries. Having a tighter orbit can induce a larger orbital angular velocity $\Omega_{\rm b}$, hence resulting in a stronger resonant interaction (cf. Eq.\,\eqref{eqs.14}) for a given CBD disk. $2.75\,\mathit M_{\odot}$ + BH has a shorter orbital separation compared to other remnants (Sect.\,\ref{post-CE_method}), therefore expressing a stronger resonant interacting for a given CBD. Consequently, more He star + BH mergers occur in the CBDs models in the case of $2.75\,\mathit M_{\odot}$ + BH binary.

In the case of BH companion, the CBD-binary interactions may help in facilitating the GW mergers for the post-CE binaries, which is contradictory to the effect of binary MT. For the post-CE binary with a remnant RSG of $2.75\,\mathit M_{\odot}$, $2.80\,\mathit M_{\odot}$ and $2.85\,\mathit M_{\odot}$, the double compact merger fraction without any interactions is around $37\%$, $16\%$ and $7\%$, respectively. The binary MT episode expands the binary orbit in the case of BH companion (Table\,\ref{tab:pre-after-MT}), hence decreasing the final merger fraction to around $21\%$, $9\%$ and $5\%$, respectively. However, the CBD-binary interactions may increase the merger fraction by a factor of $2-3$ (Fig.\,\ref{fig:disk_model_all_BH_2_75}{\color{blue}{d}}, Fig.\,\ref{fig:disk_model_all_BH}{\color{blue}{d}} and Fig.\,\ref{fig:disk_model_all_BH_2_85}{\color{blue}{d}}).

\subsection{NS companion cases}

\begin{figure}
\begin{center}
\includegraphics[width=0.5\textwidth]{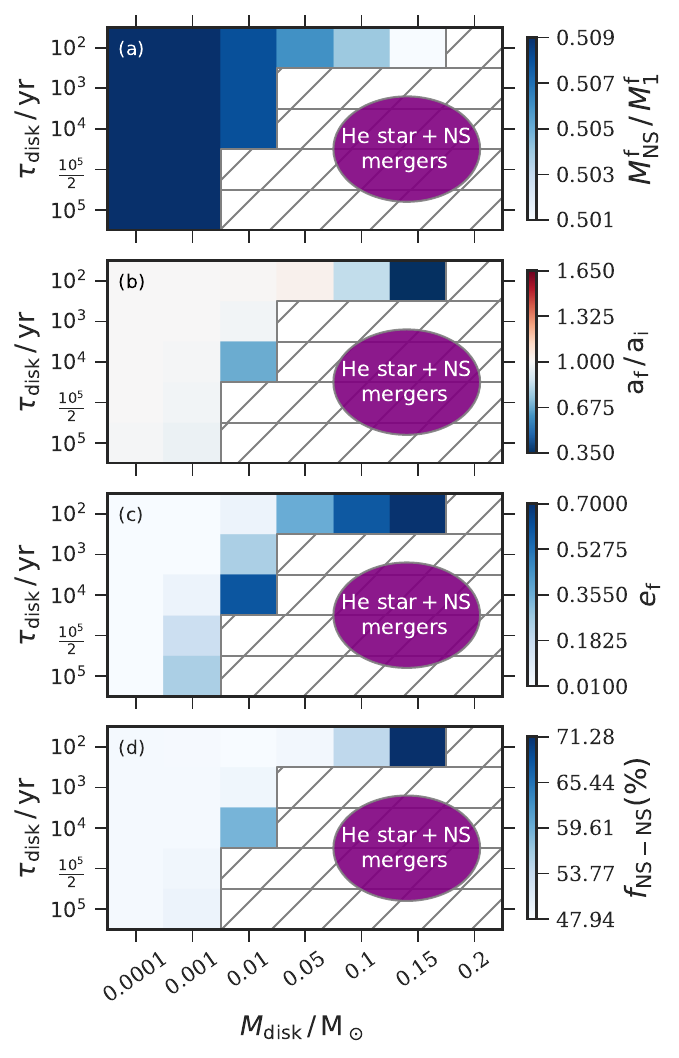}
\caption{Same as Fig.\,\ref{fig:disk_model_all_NS} but for the $2.75\,\mathit M_{\odot}$ + NS binary systems. For comparison, the double compact merger fraction of $2.75\,\mathit M_{\odot}$ + NS without any interactions is ${\sim}\, 48\%$, and around $64\%$ of binaries merge within a Hubble time after binary MT.}
\label{fig:disk_model_all_NS_2_75}
   \end{center}
\end{figure}

\begin{figure}
\begin{center}
\includegraphics[width=0.5\textwidth]{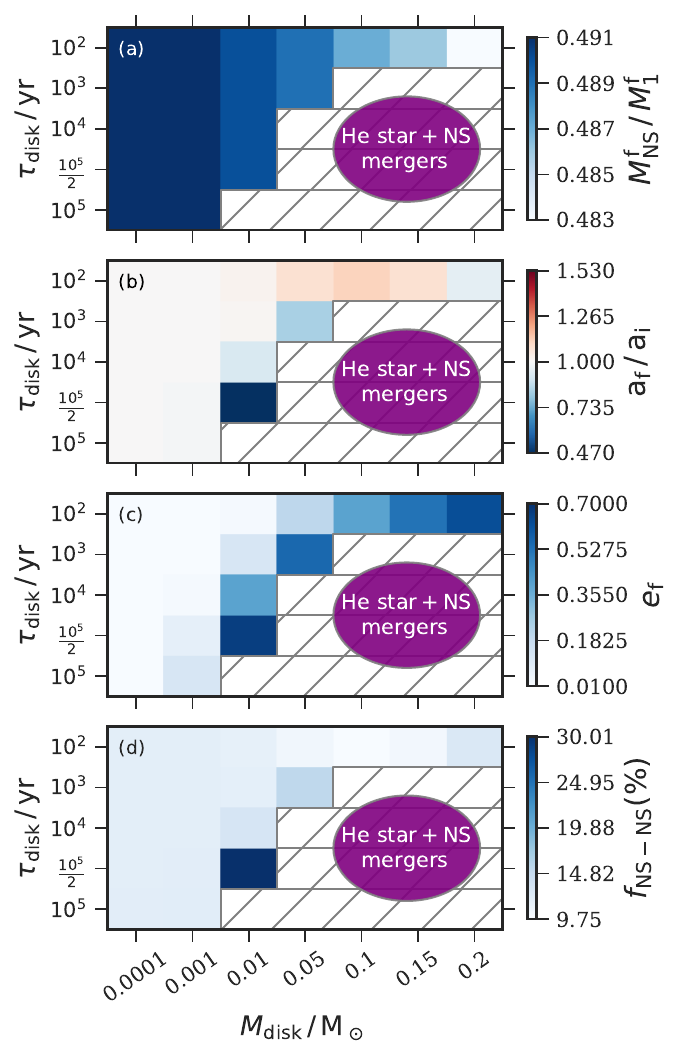}
\caption{Same as Fig.\,\ref{fig:disk_model_all_NS} but for the $2.85\,\mathit M_{\odot}$ + NS binary systems. For comparison, the double compact merger fraction of $2.85\,\mathit M_{\odot}$ + NS without any interactions is ${\sim}\, 11\%$, and around $20\%$ of binaries merge within a Hubble time after binary MT.}
\label{fig:disk_model_all_NS_2_85}
   \end{center}
\end{figure}

In the case of a NS companion, the initial orbit of the fixed binary is much tighter than the binary with a BH companion (Table\,\ref{tab:pre-after-MT}), inducing a stronger resonant interaction for a given CBD. With a shorter separation, the post-CE binary with a smaller RSG remnant can produce more He star + NS mergers, because of the stronger orbital contraction induced by the stronger resonant interaction (Fig.\,\ref{fig:disk_model_all_NS_2_75}, Fig.\,\ref{fig:disk_model_all_NS} and Fig.\,\ref{fig:disk_model_all_NS_2_85}).

The CBD-binary interactions play the same crucial role as the binary MT in the evolution of the post-CE binary. Without any interactions, the double compact merger fraction of $2.75\,\mathit M_{\odot}$ + NS, $2.80\,\mathit M_{\odot}$ + NS and $2.85\,\mathit M_{\odot}$ + NS, is around $48\%$, $25\%$ and $11\%$, respectively. Orbital shrinkage due to the binary MT can increase the merger fraction to around $64\%$, $44\%$ and $20\%$ (Table\,\ref{tab:pre-after-MT}). For the $2.75\,\mathit M_{\odot}$ + NS, $2.80\,\mathit M_{\odot}$ + NS, and $2.80\,\mathit M_{\odot}$ + NS, the CBD-binary interactions may at most increase the merger fraction to around $71\%$, $48\%$ and $30\%$, respectively.

\section{Effect of the increased eccentricity on the post-SN binary configurations}
\label{effects of increased eccentricity on the post-SN binary configurations}

\begin{figure*}
\begin{center}
\includegraphics[width=1.0\textwidth]{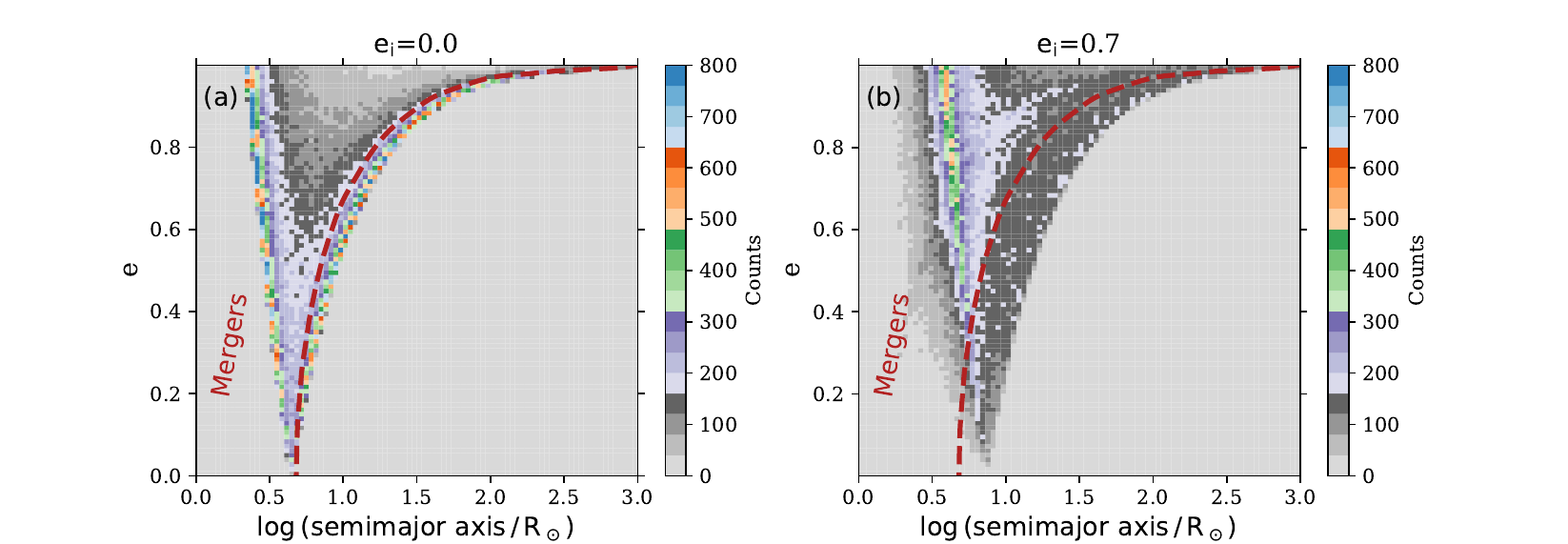}
\caption{Post-SN orbital configurations of the bound binaries with and without initial eccentricity. The models in the upper left region above the red line can merge within a Hubble time. The initial eccentricity of the binary model in panels a and b is $e_{\rm i} = 0$ and $e_{\rm i} = 0.7$, respectively.}
\label{fig:e_on_post-SN}
   \end{center}
\end{figure*}

\begin{figure*}
\begin{center}
\includegraphics[width=1.0\textwidth]{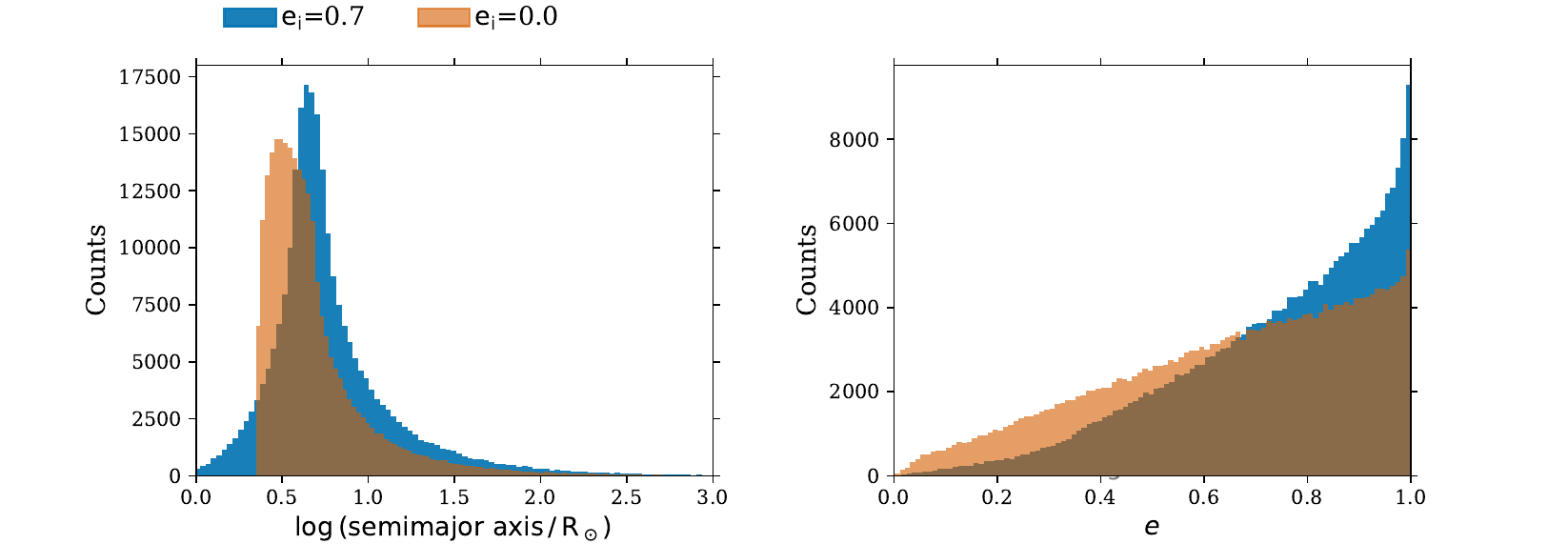}
\caption{Histogram of post-SN orbital configurations of the bound binaries which can merge in a Hubble time.}
\label{fig:e_on_post-SN_merger}
   \end{center}
\end{figure*}

SN explosions can impart kicks on the new-born NSs, affecting the binary configurations and therefore influencing the merger fraction within a Hubble time (see Sect.\,\ref{GW merger rate for MT}). To investigate the dependence of the initial eccentricity on the post-SN binary configuration, we use the fiducial model of $2.8\,\mathit M_{\odot}$ and a NS companion in an orbit of $0.53\,\rm d$ and perform $10^6$ Monte Carlo experiments of NS kicks following the kick formalism of \citet{2002ApJPfahl} to study the post-SN orbital configurations. We find that the merger fraction of binaries within a Hubble time is about $25\%$, whether in the case of the initial circular binary or eccentric binary. However, the bound binaries exhibit different distributions of eccentricity and semimajor axis after the SN explosion, as shown in Fig.\,\ref{fig:e_on_post-SN}. The initial eccentric binary tends to show higher eccentricity and larger separation after the NS explosion (Fig.\,\ref{fig:e_on_post-SN}{\color{blue}{b}}), compared to the initial circular binary (Fig.\,\ref{fig:e_on_post-SN}{\color{blue}{a}}). A similar phenomenon also exists in the bound binaries which can merge within a Hubble time as shown Fig.\,\ref{fig:e_on_post-SN_merger}. The merger fraction changes with different timescales. Within three times the Hubble time, the merger fraction of the initial circular binary is $33.0\%$, while that for the initial eccentric binary is $28.8\%$. However, within one-third of the Hubble time, the merger fractions of the initial circular binary and the eccentric binary are $19.5\%$ and $21.08\%$, respectively. In future work, we will investigate in more detail the influence of the SN kicks on an eccentric binary.

\end{appendix}

\end{document}